\documentclass[superscriptaddress,floatfix,nofootinbib,longbibliography,prx,twocolumn]{revtex4-1}
\usepackage{latexsym}
\usepackage{amssymb}
\usepackage{graphicx}
\usepackage{amsmath}
\usepackage{ulem}
\usepackage{bm}
\usepackage[colorlinks,
          linkcolor=black,
            citecolor=black,
            urlcolor=blue
           ]{hyperref}
\usepackage{verbatim}
\usepackage{mathrsfs}
\usepackage{extarrows}
\usepackage{comment}
\usepackage{mathtools,slashed}
\usepackage{soul}
\usepackage[toc,page]{appendix}
\usepackage[vcentermath]{youngtab}
\usepackage{multirow}
\usepackage{atbegshi,picture}
\usepackage{lipsum}
\usepackage{bbm}

\newcommand{\p}[1]{$\mathcal{P}(\text{Inv}_{#1}^G)$}

\newcommand{\h}{\,\text{-}\,}

\begin{document}

\title{Modulating Hamiltonian Approach to Quantum Many-Body Systems and Crystalline Topological Phases Protected by Generalized Magnetic Translations}

\author{Yuan Yao}
\email{smartyao@sjtu.edu.cn}
\affiliation{Institute of Condensed Matter Physics, School of Physics and Astronomy, Shanghai Jiao Tong University, Shanghai 200240, China}


\author{Akira Furusaki}
\affiliation{Condensed Matter Theory Laboratory, RIKEN CPR, Wako, Saitama 351-0198, Japan}
\affiliation{Quantum Matter Theory Research Team, RIKEN CEMS, Wako, Saitama 351-0198, Japan}

\date{\today}

\begin{abstract}
We discuss the topology of the parameter space of invertible phases with an onsite symmetry $G$, i.e., quantum many-body ground states that have neither fractionalization nor spontaneous breaking of the symmetry.
The classification of invertible phases is known to be obtained by counting the connected components in the parameter space of the invertible phases.
We consider its generalization --- the deformation classes of the mappings from $n$-dimensional spheres $S^n$ to this parameter space for arbitrary integer $n$.
We argue a direct one-to-one correspondence in the framework of lattice models between the non-contractibility of $S^n$ and (i) the classification of invertible phases in $d$ dimensions when $d\geq n$; or (ii) zero-dimensional invertible Hamiltonians parametrized by $S^{n-d}$ when $d<n$,
using an isotropic modulating Hamiltonian approach.
Explicitly,
we construct the noncontractible spheres of two-dimensional invertible phases,
i.e., $n=2$ and $d=2$.
We also propose a large class of crystalline topological phases protected by a generalized magnetic translations.

\end{abstract}

\maketitle
\tableofcontents

\section{Introduction}
Understanding various phases and phase transitions of many-body quantum systems is an essential topic in condensed matter physics.
Significant progress has been recently made in the studies of a large class of short-range entangled topological phases~\cite{Gu:2009aa},
and more generally, invertible topological-ordered phases,
which have a unique and gapped ground state on any lattice (manifold) without boundary~\cite{Kitaev:2011aa,Kitaev:2013aa}.
The invertible topological phases include symmetry-protected topological (SPT) phases in the presence of some onsite symmetry $G$~\cite{Gu:2009aa,Chen:2010aa,Pollmann:2012aa,Wen:TOreview2013},
as well as gapped phases without any symmetry imposed ($G=1$),
e.g., Kitaev $p$-wave superconductors~\cite{Kitaev:2001aa}.
The invertible topological phases have been investigated using various approaches
such as the K-theory classification for free electronic systems~\cite{Schnyder:2008aa,Ryu2010,Kitaev2009}, group cohomology for bosonic systems~\cite{Gu:2009aa,Pollmann:2012aa}, cobordism groups for fermionic systems~\cite{Kapustin:2014ac,Kapustin:2015aa},
and generalized cohomology~\cite{Gaiotto:2019aa}.
Two invertible phases are distinct if they cannot be connected by {any} adiabatic transformation that does not close a bulk gap as long as $G$ is preserved.
It implies the existence of an inevitable quantum phase transition between distinct invertible phases in any $G$-respecting phase diagram at zero temperature.
Such a sharp distinction of phases is indicated by a bulk topological invariant.
A characteristic property of invertible phases with nontrivial topological invariants is the presence of gapless boundary modes at spatial boundaries.
The ingappability of boundary modes can be attributed to quantum anomaly,
e.g., 't Hooft anomaly~\cite{tHooft:1980xss} associated with the symmetry $G$~\cite{Seiberg:2016aa,Witten:2016aa}, or invertible gravitational anomalies~\cite{Witten:1985aa}.

In the entire parameter space whose coordinate axes are (all possible) local interaction couplings in Hamiltonians with onsite symmetry $G$
on a $d$-dimensional lattice,
the parameter space {or \textit{the} phase diagram} of invertible systems, \p{d}, is divided into several disconnected components of distinct invertible phases partitioned by the inevitable phase transitions.
Here Inv$_d^G$ denotes the abstract set of $d$-dimensional invertible Hamiltonians with onsite symmetry $G$,
{and \p{d} is a coordinate system for Inv$_d^G$.
The classification of invertible phases is {given} by $[S^0,$\p{d}$]_0$, which is defined as the homotopy classes of mapping from the 0-sphere $S^0$ to \p{d},
and the subscript ``$_0$'' indicates the imposed condition that the ``south pole'' $x_0\in S^0$ is mapped to a Hamiltonian in the trivial phase of \p{d}.
Then the only {remaining degree} of freedom in $[S^0,$\p{d}$]_0$ is where the north pole of $S^0$ is mapped.
{Thus} $[S^0,$\p{d}$]_0$, as a set, is the same as the connected components of \p{d}.}
Later in Sec.~\ref{prep}, we will elaborate on the definition of the classification.

However,
$[S^0,$\p{d}$]_0$ only tells us about the simplest topological structure, connectivity, of the phase diagram.
The parameter space of invertible systems \p{d} can be further characterized by 
more intricate topological structures, e.g., the presence or absence of intrinsically non-contractible loops or spheres in the invertible phase diagram.
Here ``intrinsically non-contractible'' means that a certain loop or (hyper)sphere cannot be shrunk continuously to a point within \p{d}.
If such a loop is forced to contract, it necessarily touches a gapless point or a non-invertible phase in the parameter space.
The higher topological structures of \p{d} are represented by the homotopy class $[S^1,\mbox{\p{d}}]$ for the contractibility of loops (i.e., a circle $S^1$) and the $n$th homotopy class $[S^n,\mbox{\p{d}}]$ for $n$-sphere $S^{n}$ in general.
Here $[S^n,\mbox{\p{d}}]$ is defined as the maps from $S^n$ to \p{d} modulo continuous deformations.
It is also useful to introduce a smaller set $[S^n,\mbox{\p{d}}]_0$ with the additional subscript ``$[\cdots]_0$'' to specify that the maps are required to map a fixed point $\vec{\tau}_0\in S^n$ to a Hamiltonian in the trivial phase of \p{d}.
(All these notations above will be elaborated in Sec.~\ref{prep}.)

Nontrivial higher topological structures have been studied from various perspectives, e.g., defect classifications in non-interacting electronic band theories~\cite{Teo:2010aa}, 
generalized Thouless pumping~\cite{Hsin:2020aa},
and parameter-space anomaly~\cite{Cordova:2020aa,Cordova:2020ab} in some field theories.
However,
generic strongly-interacting systems such as spin models do not necessarily have a field-theory description.
In fact, lattice constructions have been employed to study the higher Berry phase~\cite{Kapustin:2020aa,Wen:2021aa}, higher U$(1)$-charge pumping~\cite{Kapustin:2020ab}, and adiabatic cycles in quantum spin models~\cite{Shiozaki:2021aa,Ohyama:2023ty} and interacting fermionic systems~\cite{Ohyama:2022ts}.
Also, the homotopy groups
$\pi_{0,1,2}$ of \p{d} in the spin-$1/2$ antiferromagnets were discussed in connection with quantum ingappabilities~\cite{Yao:2022vh}.
A suspension construction on lattices proposed earlier~\cite{Wen:2021aa} can in principle provide higher homotopy groups by several iterations of suspensions assuming an $\Omega$-spectrum structure~\cite{Kitaev:2011aa,Kitaev:2013aa}.
We note, however, that these approaches do not treat all spatial directions on equal footing, and an isotropic lattice construction should be better suited for comparison with numerical calculations and experiments.
Thus, an explicit and isotropic lattice manifestation of higher homotopy of the parameter space \p{d} still remains open.

On the other hand,
when $G$ includes lattice spatial symmetry,
the corresponding invertible phases,
i.e., crystalline topological phases~\cite{Fu:2011wr,Hsieh:2012tq,Fang:2012wp,Chiu:2013tk,Morimoto:2013wo,Essin:2013wg,Zhang:2013wh,Shiozaki:2014wd,Hsieh:2014us,Qi:2015wj,Isobe:2015tf,Cheng:2016aa,Chiu:2016aa,Hermele:2016ws,Thorngren:2018aa,Else:2019ue},
are technically more complicated to describe than the above-mentioned invertible phases where $G$ is purely onsite symmetry,
because the effects of the crystalline symmetry cannot always be captured by some continuum field theory in general dimensions.
Similarly to the invertible phases with onsite symmetry,
the boundary modes of nontrivial crystalline topological phases are forbidden to be gapped out,
as long as the bulk symmetry is not explicitly broken at the boundary.
However,
the key difference from the onsite-symmetric case is that the boundary Hamiltonian for such boundary modes can be realized in its own dimension(s) without the bulk, since the associated anomaly only implies the impossibility of realizing the bulk symmetry (which is already non-onsite) as an onsite symmetry on the boundary. 
The famous Lieb-Schultz-Mattis (LSM) theorem~\cite{Lieb:1961aa} and its various extensions~\cite{Affleck:1986aa,Oshikawa:2000aa,Hastings:2004ab,Po:2017aa,Shiozaki:2022wm,Ogata:2018aa,Else:2020aa,Ogata:2020aa} can be understood as a boundary realization of some nontrivial crystalline topological phase~\cite{Cheng:2016aa,Furuya:2017aa,Yao:2019aa}.
A natural interesting question is whether the higher topological structures of invertible phases with \textit{onsite} symmetry have intrinsic relations with crystalline topological phases.
Such connections would be useful in constructing unknown lattice crystalline phases from more familiar invertible phases with onsite symmetry.
Furthermore,
such constructions can produce generalizations of LSM-type theorems
when we focus on their boundaries. 

In this work,
we use a modulating Hamiltonian approach to argue a direct one-to-one correspondence between {$[S^n,$\p{d}$]$} and (i) the direct product {$[S^0,$\p{d-n}$]_0\times [S^0,$\p{d}$]_0$} when $d\geq n$, or
(ii) {$[S^{n-d},$\p{0}$]_0\times [S^0,$\p{d}$]_0$} when $d<n$.
To illustrate the underlying idea,
we first consider the contraction of loops and spheres in \p{d} before discussing general cases.
In general, the construction is model-dependent and requires an additional assumption.
Nevertheless,
we explicitly give a \textit{model-independent} lattice construction of $[S^2,\mbox{\p{2}}]_0$, as a special case of $d=n=2$.
Furthermore,
we apply the modulating Hamiltonian method to propose a large class of crystalline topological phases protected by a generalized magnetic translation.
This translation is a combination of the lattice translation and a symmetry that can transform a topological phase to its inverse.
Our result is general and can reproduce the original LSM theorem~\cite{Lieb:1961aa} and give its magnetic translation generalizations.

\section{Notations and preparations}\label{prep}
Except in Sec.~\ref{crystalline},
we will always assume that all the Hamiltonians preserve an onsite symmetry denoted by $G$.
In this part,
several notations related to invertible topological phases will be introduced.
We denote by $I_{d}^\text{tr}$, where ``tr'' stands for ``trivial'', a point in the $d$-dimensional gapped phase in the parameter space,
whose corresponding lattice Hamiltonian
$E_d=\mathcal{H}(I_d^\text{tr})$ is ``co-1'' (defined below) atomic.
Here $\mathcal{H}$ maps parameters to the corresponding lattice Hamiltonian.

\subsection{Invertible phases}
\begin{figure}[t]
\centering
\includegraphics[width=8.8cm,pagebox=cropbox,clip]{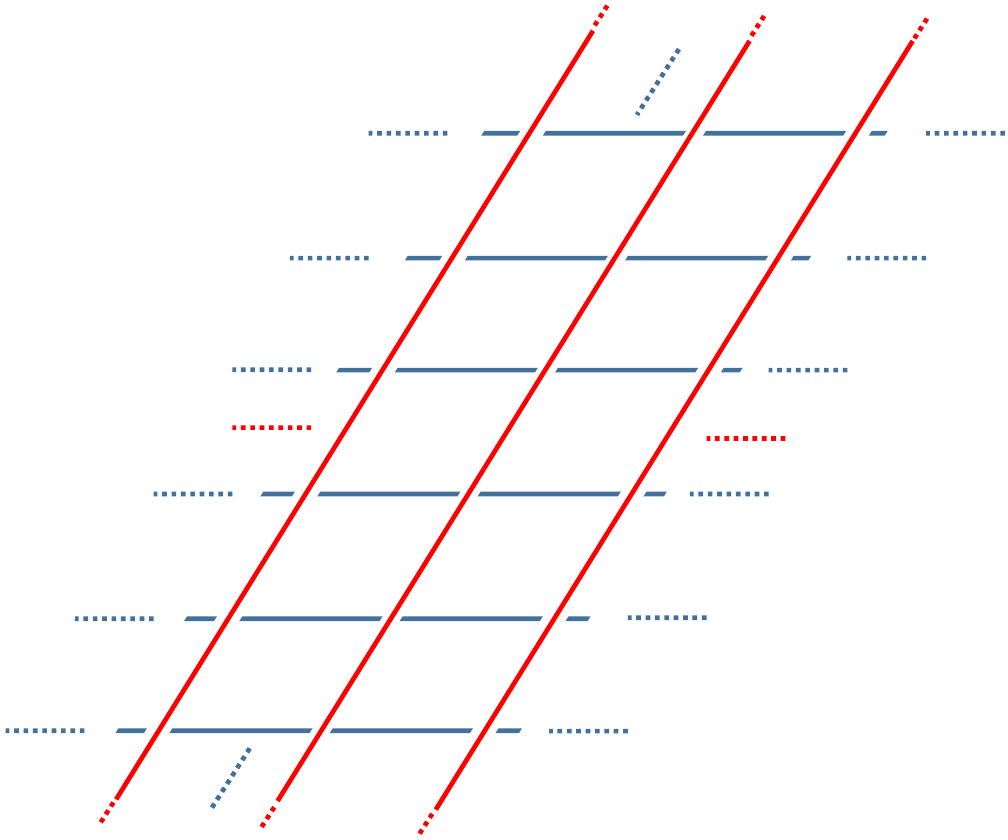}
\caption{A typical case of co-$1$ Hamiltonian in $d=2$: it is a stack of two (red and blue) two-dimensional Hamiltonians, each of which is an infinite stacking of one dimensional systems (lines).}
\label{co_1}
\end{figure}

We first define the concept of ``co-1'' atomic Hamiltonians in $d$-dimensional space:
a $d$-dimensional Hamiltonian $E_d$ is co-1 atomic if and only if it can be written as a finite stacking of $d$-dimensional gapped Hamiltonians each of which is an infinite stacking, along one of the $d$ dimensions, of decoupled $(d-1)$-dimensional gapped Hamiltonians.
A typical case of co-$1$ Hamiltonian in $d=2$ is shown in FIG.~\ref{co_1}.

More generally,
we can also define ``co-$k$'' atomic Hamiltonian in $d$ dimensions (with $k\leq d$) as \textit{a finite stacking} of $d$-dimensional Hamiltonians each of which is a stacking of $(d-k)$-dimensional gapped Hamiltonians along $k$ of $d$ dimensions.
By this definition,
co-$m$ atomic Hamiltonian is also co-$n$ atomic if $m\geq n$.
Incidentally, the conventional atomic insulators in $d$ dimensions are just co-$d$ atomic Hamiltonians.
In this work,
``an atomic Hamiltonian'' without an additional modifier will always implicitly imply a co-$1$ atomic Hamiltonian to simplify the notations.

(One more additional technical definition is that co-$1$ atomic zero-dimensional Hamiltonian is defined as a zero-dimensional Hamiltonian with trivial symmetry charges, if any, and trivial fermion parity.)

Then we can define the concept of ``being in the same phase'' as follows:
two $d$-dimensional lattice invertible Hamiltonians $H_1$ and $H_2$ are in the same phase,
if and only if
there exist two co-1 atomic $E_d$ and $E_d'$ such that there is an adiabatic path connecting the stacking of $H_1$ and $E_d$ and the stacking of $H_2$ and $E_d'$,
denoted as
\begin{eqnarray}\label{def}
H_1\h E_d  \Leftrightarrow H_2\h E_d',
\end{eqnarray}
where ``$\h$'' is the stacking operation, and
$\Leftrightarrow$ represents the existence of an adiabatic path connecting the stacked Hamiltonians on the left- and right-hand sides.
The ``adiabatic path'' in this work always means a path of Hamiltonians on which the gap is never closed.
$E_d$ and $E_d'$ are used to denote some co-1 atomic Hamiltonians.
The prime of $E_d'$ just implies that it is not necessarily the same as $E_d$.
We will use double primes if a single prime has been already used in the same formula.

More importantly,
we will call a $d$-dimensional Hamiltonian $H$ to be trivial if it is in the same phase as some co-1 atomic Hamiltonian $E_d$.

After the parameters of any $H_1$ and $H_2$ in the same phase are identified,
the parameter space \p{d} will be reduced to the set (or equivalence class) called the classification of $d$-dimensional invertible phase denoted by $[S^0,\text{\p{d}}]_0$. 

Naively, it seems more natural to define the equivalence~\eqref{def} without $E_d$ or $E'_d$,
namely to regard $H_1$ and $H_2$ as in the same phase if $H_1 \Leftrightarrow H_2$.
However, we propose Eq.~\eqref{def} as a more appropriate definition, for the following reasons:
\begin{enumerate}
\item $H_1$ and $H_2$ can have different lattice structure,
so their underlying local Hilbert space can be significantly different.
$E_d$ and $E_d'$ are supposed to balance this discrepancy;
\item Even if $H_1$ and $H_2$ have the same local Hilbert space,
they can belong to different topological phases when put on a finite-size lattice and seen as lower dimensional systems (see two examples below);
\item The global onsite symmetry $G$ of $H_1$ and $H_2$ may have a normal subgroup $K$ that can be directly taken as a gauge symmetry (see an example below), such that $H_1$ and $H_2$ belong to distinct phases under symmetry $G/K$.
Then without $E_d$ and $E_d'$,
there cannot be any $G$-symmetric adiabatic path between $H_1$ and $H_2$ even though they are in the same phase defined as in Eq.~(\ref{def}).
\end{enumerate}

The first reason above is natural and does not need further explanation.
A typical example related to the second reason is the case of $G=\mbox{U(1)}$ and $d=2$,
where one of the relevant topological invariants is the electric Hall conductance.
For example,
two fermionic models on the square lattice can have the same electric Hall conductance even if their charge densities are different (since the electric Hall conductance is determined by the ratio of charge density and the magnetic-flux density in the continuous space and up to an ambiguity on lattice~\cite{Lu:2017aa,Yao:2021aa}).
Therefore,
when put on a finite square lattice,
there cannot be any U(1)-symmetric adiabatic path between these two Hamiltonians due to their different total U(1) charges.
In other words,
they belong to distinct U(1)-symmetric topological phases when viewed as zero-dimensional systems.
The role of $E_d$ and $E_d'$ above is to fill in this difference of these zero-dimensional topological invariants.
Another less obvious example is a three-dimensional co-1 atomic insulator schematically denoted as
\begin{eqnarray}\label{atom}
\cdots\h\{\sigma_\text{H}=1\}\h\{\sigma_\text{H}=1\}\h\{\sigma_\text{H}=1\}\h\cdots,
\end{eqnarray}
which is constructed by infinitely stacking two-dimensional integer quantum Hall slices with $\sigma_\text{H}=1$.
Its ground state is long-range entangled and cannot be transformed to a conventional (co-$d$) atomic state.
However,
this nontrivial property comes from its lower dimensional slices and encoded in $[S^0,\text{\p{2}}]_0$;
regarding (\ref{atom}) as nontrivial would result in over-counting of nontriviality in $[S^0,\text{\p{3}}]_0$~\cite{Shirley:2018tg}.
Another way to see that it is reasonable to adopt the definition~(\ref{def}) such that Eq.~(\ref{atom}) is trivial is as follows.
The co-1 atomic insulator (\ref{atom}) can be adiabatically transformed to
\begin{eqnarray}
\cdots\h\{\sigma_\text{H}=0\}\h\{\sigma_\text{H}=2\}\h\{\sigma_\text{H}=0\}\h\{\sigma_\text{H}=2\}\h\cdots,
\end{eqnarray}
and further to
\begin{eqnarray}
\cdots&&\h\{\sigma_\text{H}=0\}\h\{\sigma_\text{H}=0\}\h\{\sigma_\text{H}=3\}\h\nonumber\\
&&\h\{\sigma_\text{H}=0\}\h\{\sigma_\text{H}=0\}\h\{\sigma_\text{H}=3\}\h\cdots,
\end{eqnarray}
by ``accumulating'' the quantum Hall conductance.
We can repeat this accumulation procedure arbitrarily further,
so the resultant system is $\sigma_\text{H}=0$ almost everywhere (thus trivial), except on very diluted slices with large $\sigma_\text{H}$.
With this intuitive picture,
we can also remark the importance of $G$ to be purely onsite symmetry,
e.g., if $G$ contains translation symmetry along the stacking direction,
the above accumulations break $G$, and indeed Eq.~(\ref{atom}) is a nontrivial weak topological phase~\cite{Fu:2007aa}.

According to the definition~\eqref{def}, the co-1 atomic insulator~\eqref{atom}
is trivial \textit{as a $3$-dimensional phase}.
Moreover, the Hall conductance can be eliminated by stacking \eqref{atom} with the co-1 atomic insulator built of 
$\sigma_\text{H}= -1$ layers that belongs to the same trivial phase.
Of course, a $\sigma_\text{H} \neq 0$ layer is nontrivial as a $2$-dimensional phase,
since it cannot be trivialized by stacking with a co-1 atomic insulator in $2$ dimensions.

From the definitions explained above,
we can obtain the following useful corollary for any co-$1$ atomic $E_d$:
\begin{eqnarray}\label{useful}
E_d\h E'_d\Leftrightarrow E_d'',
\end{eqnarray}
where $E'_d$ is some co-1 atomic Hamiltonian while $E_d''$ is a conventional co-$d$ atomic Hamiltonian.
Intuitively,
Eq.~(\ref{useful}) can be understood as trivializing the lower-dimensional topological phases inside $E_d$, dimension by dimension, down to zero dimension.
Considering the intermediate steps of such a trivialization,
we can make a stronger statement where $E''_d$ in Eq.~(\ref{useful}) can be co-$(d-k)$ atomic for arbitrary $k<d$,
but we will not use this generalization.

We can also show that two arbitrary co-$1$ atomic Hamiltonians $E_d'$ and $E_d''$ share some ``terminal'' conventional atomic Hamiltonian $E_d$ in the following sense:
for arbitrary co-$1$ $E_d'$ and co-$1$ $E_d''$,
there exists a conventional co-$d$ atomic $E_d$,
co-$1$ $E_d'''$ and co-$1$ $E_d''''$ such that
\begin{eqnarray}\label{useful_1}
E_d'\h E_d'''\Leftrightarrow E_d \Leftrightarrow E_d''\h E_d''''.
\end{eqnarray}
This can be proven by using Eq.~(\ref{useful}) and the fact that the stacking of co-$m$ and co-$n$ Hamiltonians is co-min$\{m,n\}$ atomic.

The remaining third reason is much more nontrivial than the previous two.
Let us consider an SO(3)-symmetric spin-1/2 chain with a unique gapped ground state.
We do not treat this $\text{SO}(3)=\text{SU}(2)/\mathbb{Z}_2$ symmetry as SU(2), since the $\mathbb{Z}_2$ sector in SU(2) behaves as a gauge symmetry; the action of $\mathbb{Z}_2$ is only to multiply the local Hilbert space by the $\mathbb{Z}_2$ phase factor ($\pm$1), but two quantum states describe the same physical state if they differ only by a phase factor.
Nevertheless,
it is also possible to ignore this gauge structure and consider the global symmetry as SU(2).
This amounts to regarding our lattice spin model as a low-energy effective theory whose parent ultra-violet theory like a Hubbard-type model does not have the $\mathbb{Z}_2$-gauge structure
(and it is also related to the concept of symmetry extension~\cite{Witten:2016ab,Seiberg:2016aa,Wang:2018um}).
However, the group-cohomology classification~\cite{Chen-Gu-Wen_classification2010,Chen:2013aa} tells us that the one-dimensional SU(2) SPT classification is always trivial ($\{0\}$) while SO(3) SPT classification is nontrivial ($\{0,1\}=\mathbb{Z}_2$).
If we did not include $E_d$ and $E_d'$ in the definition above when taking $G=\text{SU}(2)$,
the resultant classification of invertible phases would be insensitive to $G$ being whether SU(2) or SO(3),
inconsistently with the group-cohomology classification.
Therefore,
the essential role of $E_d$ and $E_d'$ is to break the $\mathbb{Z}_2$ gauge structure of the lattice models $H_{1,2}$; a concrete simple example is presented in Appendices.
In general,
the necessity of $E_d$ and $E_d'$ is to make our definition~(\ref{def}) consistent with the group-cohomology approach that studies ground-state wavefunctions.

Let ``$A_d$'' be an element in Inv$_d^G$.
The invertibility of the lattice Hamiltonian $A_d$ implies the existence of its inverse $\bar{A}_d$;
stacking of $A_d$ and $\bar{A}_d$ is in the same phase as a co-$d$ atomic $E_d$.
By the definition~(\ref{def}),
\begin{eqnarray}
A_d\h\bar{A}_d\h E_d''\Leftrightarrow E_d\h E_d',
\end{eqnarray}
Furthermore,
we can further stack an atomic Hamiltonian $E_d'''$ on $E_d'$ such that all their lower-dimensional (except 0-dimensional) topological invariants are trivialized by Eq.~(\ref{useful}).
Namely,
$E_d'\h E'''_d\Leftrightarrow E_d$
for a conventional co-$d$ atomic $E_d$.
After redefining $(\bar{A}_d)_\text{new}\equiv \bar{A}_d\h E''_d\h E'''_d$,
we have
\begin{eqnarray}\label{inverse}
A_d\h\bar{A}_d\Leftrightarrow E_d\h E_d.
\end{eqnarray}
In particular,
$A_d\h\bar{A}_d$ and $E_d\h E_d$ now have the same local Hilbert space and lower-dimensional topological invariants.
Moreover, the classification $[S^0,\text{\p{d}}]_0$ forms an abelian group under the group addition defined by stacking ``$\h$''.

\subsection{Mappings from hyperspheres $S^n$ to parameter space \p{d}}
To characterize the non-contractible loops in \p{d},
we define equivalence relations and addition operation for loops.
A loop is a mapping $J:S^1\rightarrow\mbox{\p{d}}$ from the unit circle $S^1\equiv[0,2\pi]/\{0\sim2\pi\}$ to the parameter space \p{d}.
In other words,
$J(\tau)$ for each $\tau\in[0,2\pi]$ is the parameter corresponding to the lattice Hamiltonian $\mathcal{H}[J(\tau)]$.
In the following discussion, we will often denote this lattice Hamiltonian by $J(\tau)$ for brevity.

We may deform $J$ continuously to obtain a nearby loop $J'$ if there exists a continuous deformation $F:[0,1]\times S^1\rightarrow\mbox{\p{d}}$ such that $F(0,\tau)=J(\tau)$ and $F(1,\tau)=J'(\tau)$.
We denote this situation by
\begin{eqnarray}
J\Leftrightarrow_1 J',
\end{eqnarray}
where the subscript ``$_1$'' coming from the superscript of $S^1$ is introduced to remind us that we are dealing with loops rather than points in \p{d} as in Eq.~(\ref{def}).
More generally,
suppose that $J$ and $J'$ are mappings $S^n\rightarrow\mbox{\p{d}}$, and that there exists $F:[0,1]\times S^n\rightarrow$\p{d} with $F(0,\vec{\tau})=J(\vec{\tau})$ and $F(1,\vec{\tau})=J'(\vec{\tau})$,
then we write
\begin{eqnarray}
J\Leftrightarrow_nJ'.
\label{J Leftrightarrow_n J'}
\end{eqnarray}
Here our convention is that $\vec{\tau}$ is used to denote a point on the unit sphere $S^n$.
Thus,
$L_H\vec{\tau}$ is a point on a sphere $S^n$ with radius $L_H$.

Furthermore,
there is a stacking operation induced from the stacking ``$\h$'' of Hamiltonians in Eq.~(\ref{def}).
We will use the symbol ``+'' for this operation:
\begin{eqnarray}
J+J':S^1\rightarrow\text{\p{d} with }(J+J')(\tau)\equiv J(\tau)\h J'(\tau),\nonumber
\end{eqnarray}
where $J(\tau)\h J'(\tau)$ is meant for the stacked Hamiltonian $\mathcal{H}[J(\tau)]\h\mathcal{H}[J(\tau')]$.

There is a trivial loop $J_E$ with $J_E(\tau)=I^\text{tr}_d$ for any $\tau$; $J_E$ maps the entire circle $S^1$ to a fixed atomic Hamiltonian parameter $I^\text{tr}_d$.
Now we can generalize the concept of ``being in the same phase'' to these loops as follows:
two mappings $J$ and $J'$ from $S^1$ to \p{d} are in the same $1$-phase if and only if there exist $J_E$ and $J_{E'}$
\begin{eqnarray}\label{1_phase}
J+J_E\Leftrightarrow_1 J'+J_{E'},
\end{eqnarray}
where $\mathcal{H}[J_E(\tau)]=E_d$ and $\mathcal{H}[J_{E'}(\tau)]=E'_d$.

Such a definition is directly generalized to the mappings from $S^n$ to \p{d} by replacing all the ``1''s to ``$n\geq1$'':
\begin{equation}\label{addition}
J+J':S^n\rightarrow\text{\p{d} with }(J+J')(\vec{\tau})\equiv J(\vec{\tau})\h J'(\vec{\tau}),
\end{equation}
for $J$ and $J'$ mappings $S^n\rightarrow\mbox{\p{d}}$,
and $J$ and $J'$ are in the same $n$-phase if and only if there exist constant mappings $J_E$ and $J_{E'}$ so that
\begin{equation}\label{n-phase}
J+J_E\Leftrightarrow_n J'+J_{E'}.
\end{equation}
When a mapping $J$ is in the same $n$-phase as $J_E$,
we will call $J$ ``contractible''.

We will identify two mappings if they are in the same $n$-phase and denote the set of distinct mappings (equivalence classes) as $[S^n,\text{\p{d}}]$ with $n\geq1$.
More importantly,
$[S^n,\text{\p{d}}]$ becomes an abelian group when we take the group addition as the ``+'' defined in Eq.~(\ref{addition}).

For later convenience,
we can impose an additional condition on $J$ such that $J(\vec{\tau}_0)$ is a Hamiltonian parameter in the trivial topological phase with $\vec{\tau}_0$ the south pole of $S^n$.
The restricted classification will have an additional subscript ``$_0$'': $[S^n,\text{\p{d}}]_0$.
Actually, since $S^n$ is a connected manifold when $n\geq1$,
the image of such restricted mappings will be also in the trivial phase.
Together with the previous notation of $[S^0,\text{\p{d}}]_0$,
we have completed the definition of $[S^n,\text{\p{d}}]_0$ for any non-negative integer $n$.
Similarly,
$[S^n,\text{\p{d}}]_0$ is also an abelian group under the addition defined in Eq.~(\ref{addition}).

\begin{figure}[t]
\centering
\includegraphics[width=8.8cm,pagebox=cropbox,clip]{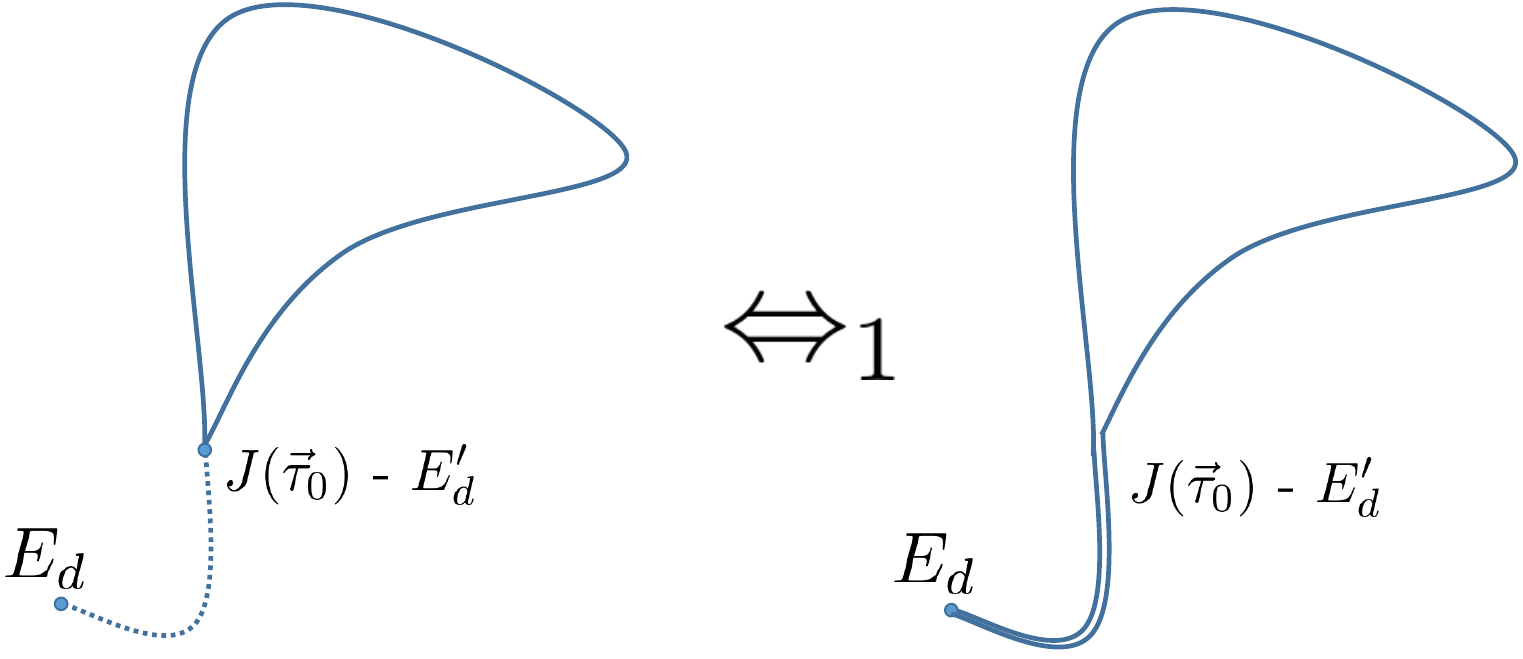}
\caption{We can pinch the south pole of the loop $(J+J_{E'})(\vec{\tau}_0)=J(\vec{\tau}_0)\h E_d'$ and stretch it to $E_d$ by the known path $J(\vec{\tau}_0)\h E_d'\Leftrightarrow E_d$. Then the loop after the deformation has its south pole mapped to $E_d$. This type of deformation can be naturally generalized to the mappings from hyperspheres $S^n$.}
\label{pinch}
\end{figure}

Another simplification can be made for
$J:S^n\rightarrow\mbox{\p{d}}$ with $J(\vec{\tau}_0)$ in the trivial atomic phase,
i.e., a representative of an element in $[S^n,\text{\p{d}}]_0$.
Using Eq.~(\ref{useful}),
we can trivialize all the lower-dimensional topological invariants of $J(\vec{\tau}_0)$ by some atomic $E_d'$ as $J(\vec{\tau}_0)\h E_d'\Leftrightarrow E_d$ to some conventional co-$d$ atomic $E_d$.
Thus,
this adiabatic path induces the following deformation shown in FIG.~\ref{pinch}:
\begin{eqnarray}\label{unity}
J+J_{E'}\Leftrightarrow_n J',
\end{eqnarray}
with $J'(\vec{\tau}_0)=E_d$, a conventional co-$d$ atomic insulator.

Based on Eqs.~(\ref{useful}) and (\ref{useful_1}),
another useful simplification can be made for arbitrary $J$ and $J'$ which are in the same $n$-phase of $[S^n,\text{\p{d}}]_0$,
i.e., satisfying
\begin{eqnarray}\label{n-phase-useful}
J+J_E\Leftrightarrow_n J'+J_{E'}
\end{eqnarray}
and $J(\vec{\tau}_0)\Leftrightarrow J'(\vec{\tau}_0)$.
Under these conditions,
we can also take $J(\vec{\tau}_0)=J'(\vec{\tau}_0)$ without loss of generality by employing ``pinching'' deformation similar to the one in FIG.~\ref{pinch},
and denote $\mathcal{H}[J(\vec{\tau}_0)]$ by $E''_d$.
Then,
by considering the two sides of Eq.~(\ref{n-phase-useful}) evaluated at $\vec{\tau}_0$,
we have
\begin{eqnarray}\label{conn}
E_d''\h E_d\Leftrightarrow E''_d\h E_d'.
\end{eqnarray}
We note that
\begin{eqnarray}\label{conn_0}
J_{E''}+J_E=J_{E''\h E},
\qquad
J_{E''}+J_{E'}=J_{E''\h E'}
\end{eqnarray}
by definition.
From Eq.~(\ref{conn}),
we obtain
\begin{eqnarray}\label{conn_1}
J_{E''}+J_E \Leftrightarrow_n J_{E''}+J_{E'}.
\end{eqnarray}
We can add $J_{E''}$ on both sides of Eq.~(\ref{n-phase-useful}) and use Eqs.~(\ref{conn_0}) and (\ref{conn_1}) to obtain
\begin{eqnarray}\label{conn_2}
J+J_{E''\h E}\Leftrightarrow_n J'+J_{E''\h E}
\end{eqnarray}
It follows from Eq.~(\ref{useful_1}) that 
$E''\h E$ can be further reduced to a conventional co-$d$ atomic insulator $E'''_d$ by stacking some $E''''_d$,
we obtain the following simplification (after adding $J_{E''''}$ on both sides of Eq.~(\ref{conn_2}) and renaming $E'''_d$ to be $E_d$): 
\begin{eqnarray}\label{useful-n-phase}
J+J_E\Leftrightarrow_n J'+J_E,
\end{eqnarray}
{with} some $J_E$ where $E_d=\mathcal{H}[J_E(\vec{\tau})]$ is a conventional co-$d$ atomic insulator.

\section{Non-contractibility of loops {$[S^1,\cdot\,]_0$}}
\begin{figure}[t]
\centering
\includegraphics[width=8.8cm,pagebox=cropbox,clip]{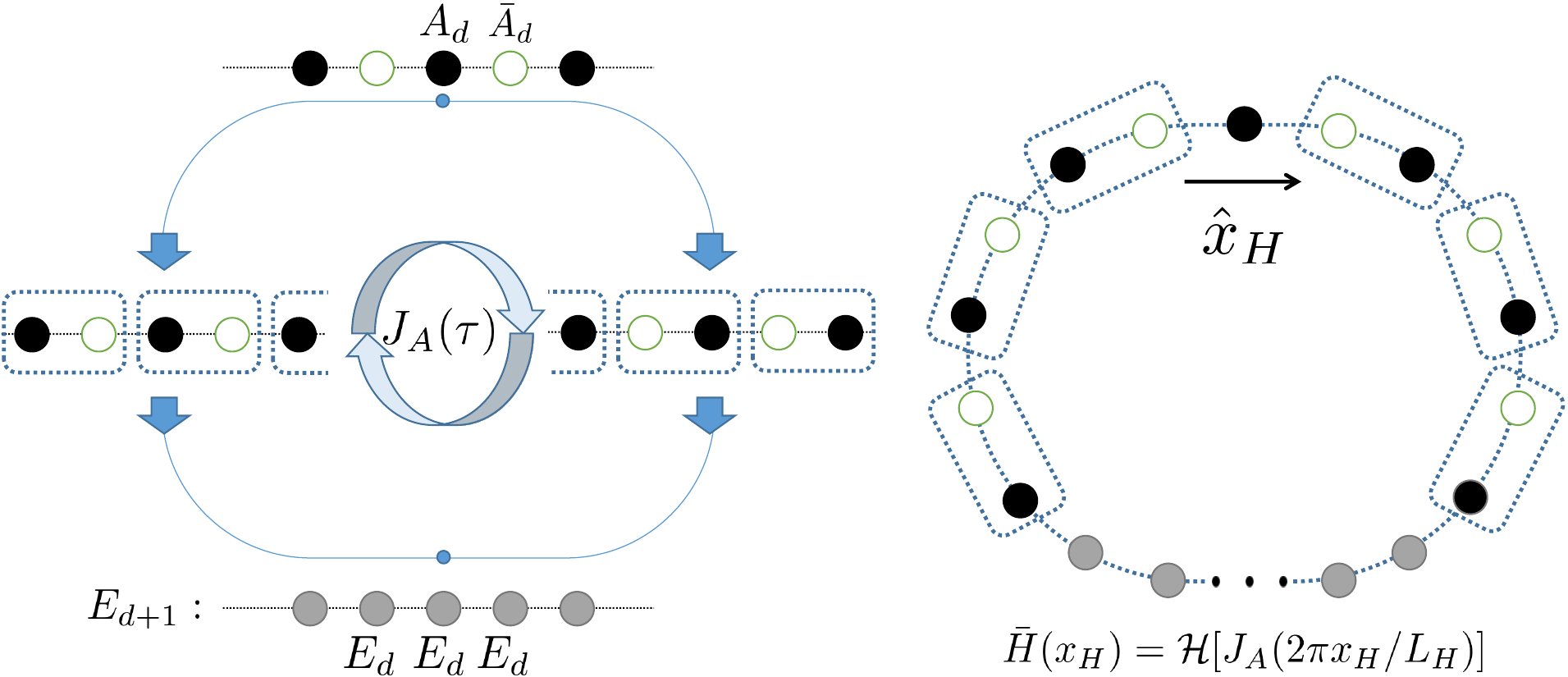}
\caption{Construction of the loop $J_A(\tau)$ (left) and the modulating Hamiltonian $\tilde{H}$ obtained from $J_A(\tau)$ (right).
{Left: the Hamiltonian on the top {(\ref{-A_d-barA_d-})} is connected,
within \p{d}, to the Hamiltonian on the bottom {(\ref{-E_d-E_d-}) via the} two paths indicated by the vertical arrows.
These two paths {are combined to} form a loop $J_A:S^1\rightarrow $\p{d} oriented by the arrows in the middle.
Right: the modulation is along the real-space direction $\hat{x}_H$ along the arrow.}
}
\label{loop_mod}
\end{figure}

We begin with the discussion of the non-contractibility of loops in the parameter space $\mathcal{P}(\text{Inv}_{d+1}^G)$ of $(d+1)$-dimensional invertible phases.
{Our discussion corresponds to the domain-wall approach in Refs.~\cite{Kitaev:2011aa,Xiong:2018aa,Gaiotto:2019aa}.}
We will first prove an inequality:
\begin{eqnarray}\label{pi1}
{[S^1,\text{\p{d+1}}]_0\geq [S^0,\text{\p{d}}]_0.}
\end{eqnarray}
Let ``$A_d$'' be an element in Inv$_d^G$.
After the procedure above Eq.~(\ref{inverse}),
stacking $A_d$ and $\bar{A}_d$ together enables an adiabatic transformation of this composed system to the stacking of conventional co-$d$ atomic Hamiltonians in the trivial phase $E_d\h E_d$.
As depicted in the left figure (top and bottom) in FIG.~\ref{loop_mod},
we build two Hamiltonians in Inv$_{d+1}^G$ by arranging elements of Ind$_d^G$ along the $(d+1)$-th spatial direction $\hat{x}_H$ $\text{($H$: ``horizontal'')}$:
a staggering array of $A_d$ and $\bar{A}_d$
\begin{equation}
\cdots\h A_d\h \bar{A}_d\h A_d\h \bar{A}_d\h \cdots,
\label{-A_d-barA_d-}
\end{equation}
and a uniform array of the trivial Hamiltonian $E_d$
\begin{equation}
\cdots\h E_d\h E_d\h E_d\h E_d\h \cdots.
\label{-E_d-E_d-}
\end{equation}
The latter is obviously in the $(d+1)$-dimensional trivial phase $[E_{d+1}]$ represented by $E_{d+1}=\mathcal{H}({I_{d+1}^\text{tr}})$ where {$I^\text{tr}_{d+1}$} represents the trivial system in $d+1$ dimensions.
Here we have introduced the notation $[X]$ for the invertible phase represented by Hamiltonian $X$.
We connect these two Hamiltonians (\ref{-A_d-barA_d-}) and (\ref{-E_d-E_d-}) in Inv$_{d+1}^G$ by the following two adiabatic paths:
one path along which each pair of neighboring $(A_d\h \bar{A}_d)$ is transformed to $(E_d\h E_d)$ by adiabatically turning on the coupling between $A_d$ and $\bar{A}_d$, and the other path along which each pair $(\bar{A}_d\h {A}_d)$ is transformed to $(E_d\h E_d)$.
The combination of these two paths defines a closed loop $J_A(\tau)$, with $\tau\in[0,2\pi]$ and $J_A(0)=J_A(2\pi)={I_{d+1}^\text{tr}}$ at the south pole of $S^1$, along which the Hamiltonians $\mathcal{H}[J_A(\tau)]$ are all invertible,
i.e., this loop (modulo ``being in the same $1$-phase'') is an element in $[S^1,$\p{d+1}$]_0$.

Let us examine whether the above loop $J_A(\tau\in[0,2\pi])$ is contractible in its connected component of \p{d+1}.
To do so,
we construct an $x_H$-dependent modulating Hamiltonian (the right panel of FIG.~\ref{loop_mod}):
\begin{eqnarray}\label{modloop}
\tilde{H}_A=\sum_{x_H}\tilde{H}_A(x_H),
\,\,
\tilde{H}_A(x_H)\equiv\mathcal{H}[J_A(2\pi x_H/L_H)].
\end{eqnarray}
Viewed as a $d$-dimensional lattice model (where the $x_H$ direction is compactified, $x_H+L_H\sim x_H$), $\tilde{H}_A$ is in the same topological phase as $A_d$, since all the neighboring $A_d$ and $\bar{A}_d$ except one $A_d$ can be locally trivialized in pair~\cite{Yao:2022vh}.

Suppose that $J_A$ can be deformed to $J_E$ which is a constant mapping to the conventional co-$(d+1)$ atomic $E_{d+1}$ in FIG.~\ref{loop_mod}.
The deformation is given by $F:[0,1]\times S^1\rightarrow\mbox{\p{d+1}}$,
and the mapping $J_s:S^1\rightarrow\mbox{\p{d+1}}$ defined as $J_s(\tau)\equiv F(s,\tau)$ gives the intermediate loops with $J_0=J_A$ and $J_1=J_E$.
Then we can also define intermediate modulating Hamiltonians as $\tilde{H}(s)=\sum_{x_H}\tilde{H}(s,x_H)$ with $\tilde{H}(s,x_H)\equiv\mathcal{H}[J_s(2\pi x_H/L_H)]$, which adiabatically connects $\tilde{H}(0)=\tilde{H}_A\in[A_d]$ and $\tilde{H}(1)=\mathcal{H}(J_E)$.
Since $\mathcal{H}(J_E)$ is topologically trivial,
$[A_d]$ must also be the trivial phase.
Thus, if $A_d$ is not in the trivial phase $[E_d]$, $J_A$ is not contractible
and the (non-)contractibility of the loop $J_A(\tau)$ depends on $[A_d]$.
We can repeat the above argument without significant changes to 
prove that $J_A$ and $J_B$ are not in the same $1$-phase if $[A_d]\neq[B_d]$.
As we discuss in more detail later around Eq.~(\ref{backward}),
the $1$-phase that $J_A$ belongs to is exactly characterized and determined by $[A_d]$;
e.g., the exact ways of the local trivializations in Eq.~(\ref{inverse}) and {in} FIG.~\ref{loop_mod} are irrelevant.

With the above arguments, Eq.~(\ref{pi1}) is proved as an equality for the sets; given each element from the smaller set, 
we can assign to it a corresponding element from the larger set.
For $[S^0,$\p{d}$]_0$ to be identified as a subgroup of 
$[S^1,$\p{d+1}$]_0$ as in Eq.~(\ref{pi1}),
their group operations,
the stacking ``$\h$'' and the addition ``$+$'', respectively, should be compatible.
We can verify this as follows.
We recall that the group addition $[J_A+J_B](\tau)$ of two loops $J_A(\tau)$ and $J_B(\tau)$, which are constructed individually from two invertible phases $A_d$ and $B_d$ as in FIG.~\ref{loop_mod}, is defined as stacking $J_A(\tau)\h J_B(\tau)$ for each $\tau$.
It exactly satisfies the definition of $J_{A\h B}$,
where $J_{A\h B}$ is the loop constructed from the system obtained by stacking $A_d$ with $B_d$, i.e., replacing ``$A_d$''s in FIG.~\ref{loop_mod} by ``$A_d\h B_d$''s.
Thus,
we have proved Eq.~(\ref{pi1}) by constructing a group-monomorphism $J:A_d\mapsto J_{A}$.

\section{Non-contractibility of $n$-shperes by $[S^n,\cdot]$}
Now we generalize Eq.~(\ref{pi1}) to
\begin{eqnarray}\label{pik}
[S^{n},\text{\p{d+n}}]_0\geq [S^0,\text{\p{d}}]_0.
\end{eqnarray}
The group addition of $[S^{n},\text{\p{d+n}}]_0$ is defined through stacking of Hamiltonians, as in the $n=1$ case above.

\begin{figure}[t]
\centering
\includegraphics[width=8.8cm,pagebox=cropbox,clip]{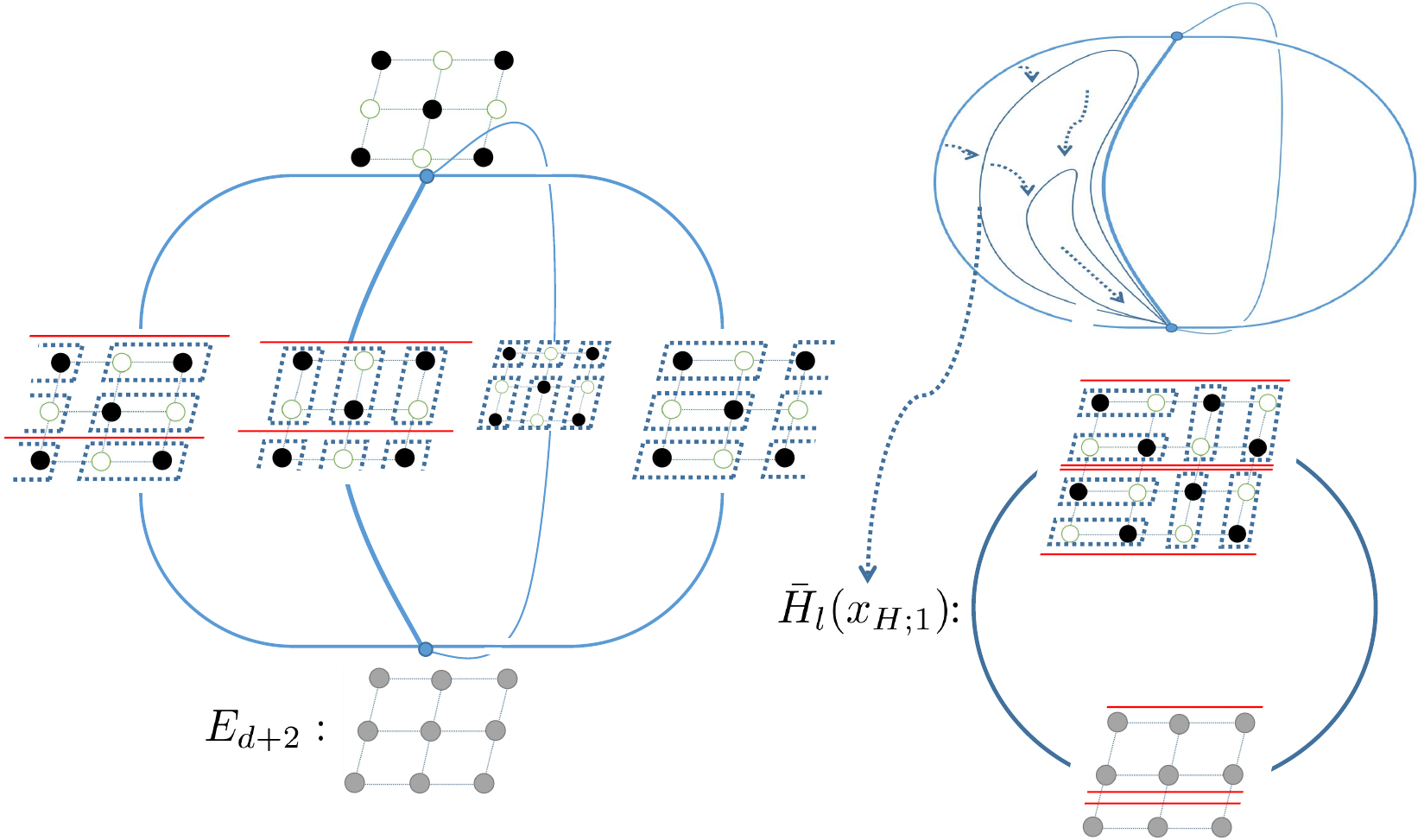}
\caption{Left: construction of the skeleton of a sphere in the parameter space.
The filled and empty circles surrounded by a dotted rectangle represents a (relatively) strongly coupled pair of $A_d$ and $\bar{A}_d$.
Right: filling of one of the four loop holes induced by the contraction of its bounding loop. 
The loop can be contracted because of the triviality of $\tilde{H}(x_{H;1})$ composed of decoupled wires of width two.}
\label{sphere}
\end{figure}

Before addressing the general $n>1$ case, we consider the case $n=2$ in the following steps for illustration.

{\bf Step 1:}
Given $A_d\in{\text{Inv}_d^G}$,
we construct a $(d+2)$-dimensional invertible system by putting $A_d$ and $\bar{A}_d$, respectively, on the two sublattices of a square lattice spanned by $\hat{x}_{H;1}$ and $\hat{x}_{H;2}$.
We also prepare a Hamiltonian in the trivial phase $[E_{d+2}]$ by putting $E_{d}$ on the above square lattice.

{\bf Step 2$_0$:}
We connect these two $(d+2)$-dimensional Hamiltonians via four different paths,
i.e., four types of pairings as shown in FIG.~\ref{sphere} (left panel), which are distinguished by the pattern of dimerization $(A_d, \bar{A}_d)$.
The four paths form four loops and can be viewed as a skeleton of a sphere.

{\bf Step 2$_1$:}
We complete the sphere
by filling in the four empty loop holes encircled by a pair of neighboring paths.
This filling procedure is possible if each boundary loop is contractible, since the contraction process naturally fills the areas.
Let us focus on one of the four loops.
Each Hamiltonian on the loop can be decomposed into decoupled wires of width two as indicated by the red lines in FIG.~\ref{sphere}.
Thus, the modulating Hamiltonian $\tilde{H}_l(x_{H;1})$, constructed from this loop by Eq.~(\ref{modloop}),
is also composed of decoupled modulating wires as illustrated in the bottom right in FIG.~\ref{sphere}.
Each decoupled width-two wire in $\tilde{H}_l(x_{H;1})$ is in the trivial phase $[E_{d+1}]$ since each $A_d$ is paired uniquely with $\bar{A}_d$ so that all of them can be locally trivialized in pairs,
in contrast to FIG.~\ref{loop_mod}.
Then all the wires can be trivialized simultaneously, so the corresponding modulating Hamiltonian is in the trivial phase,
which means that the original loop is contractible (c.f. the argument around Eq.~(\ref{backward})) after some addition (if necessary, as done in Eq.~(\ref{useful-n-phase})) of $J_{E'}$ with co-$d$ atomic $E'$.
In this way the above four loop holes can be filled completely.
The resultant surface $J_A(\vec{\tau})$ is an element in $[S^{2},\text{\p{d+2}}]_0$,
where $\vec{\tau}$ parametrizes the unit sphere.

\begin{figure}
\centering
\includegraphics[width=8.8cm,pagebox=cropbox,clip]{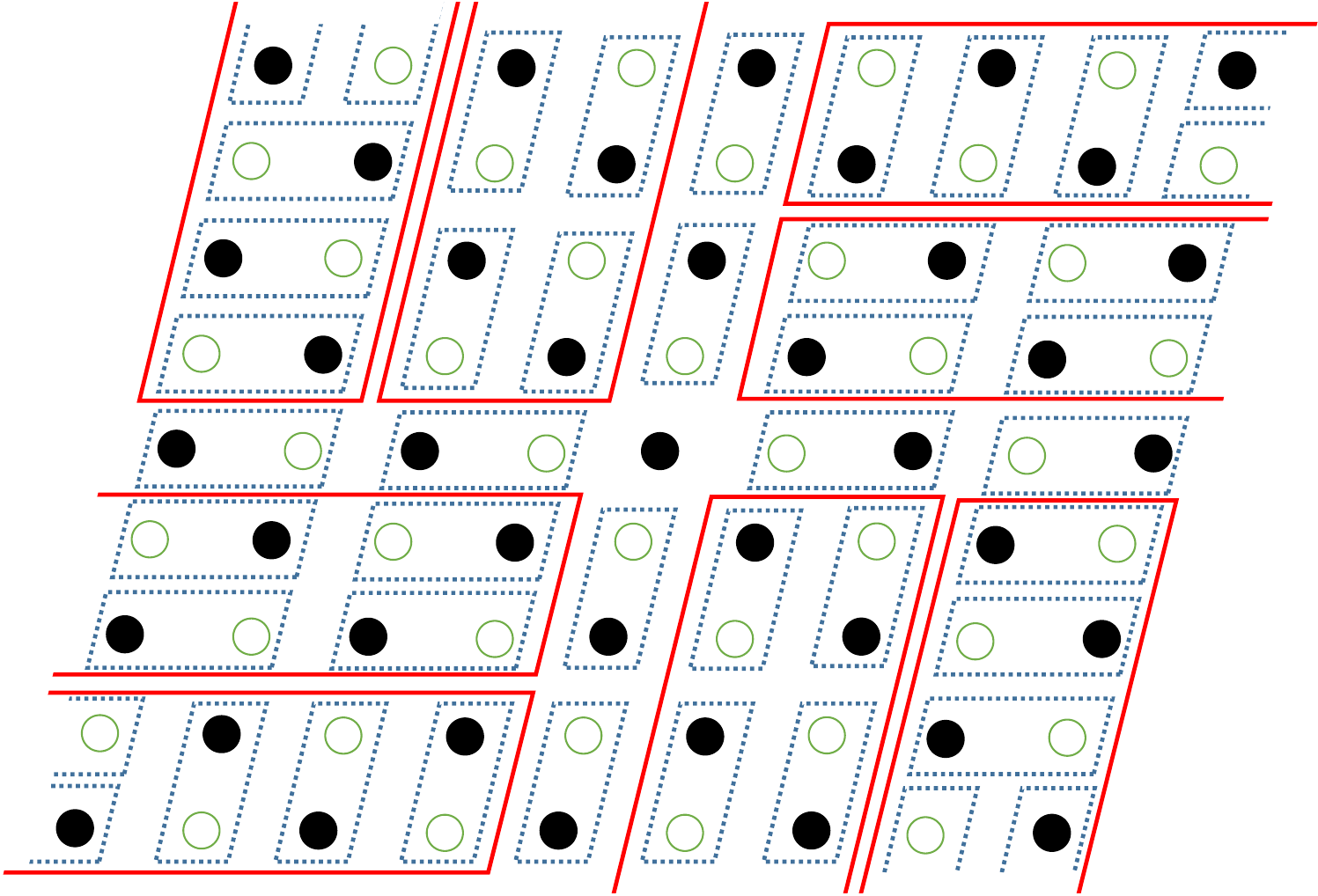}
\caption{The modulating Hamiltonian $\tilde{H}(\vec{x}_H)$ with the horizontal coordinate $\vec{x}_H=({x}_{H;1},x_{H;2})$ spanning a sphere: due to the unpaired $A_d$ at the ``north pole'',
$\tilde{H}(\vec{x}_H)$ is topologically nontrivial which implies the noncontractibility of the sphere $J_A(\vec{\tau})$ in the parameter space \p{d+2}.}
\label{modulating_sphere}
\end{figure}

{\bf Step 3:}
Finally, we examine the contractibility of the sphere $J_A(\vec{\tau})$ by taking the horizontal coordinates $\vec{x}_H=(x_{H;1},x_{H;2})$ to span a sphere of radius $L_H$.
We construct an $\vec{x}_H$-dependent modulating gapped Hamiltonian $\tilde{H}(\vec{x}_H)=\mathcal{H}[J_A(\vec{x}_{H}/L_{H})]$, which is in the same phase as the unpaired $A_d$ at the north pole of the sphere as illustrated in FIG.~\ref{modulating_sphere}; {a similar construction was previously considered for non-invertible phase diagrams~\cite{Aasen:2022aa}.}
If $A_d$ is not in the trivial phase $[E_d]$, then the sphere $J_A(\vec{\tau})$ is non-contractible, i.e., a nontrivial element of $[S^{2},\text{\p{d+2}}]_0$.
Furthermore, the same consideration as in $n=1$ allows us to conclude the compatibility of the group additions,
thereby proving Eq.~(\ref{pik}) when $n=2$.

\begin{figure}[t]
\centering
\includegraphics[width=8cm,pagebox=cropbox,clip]{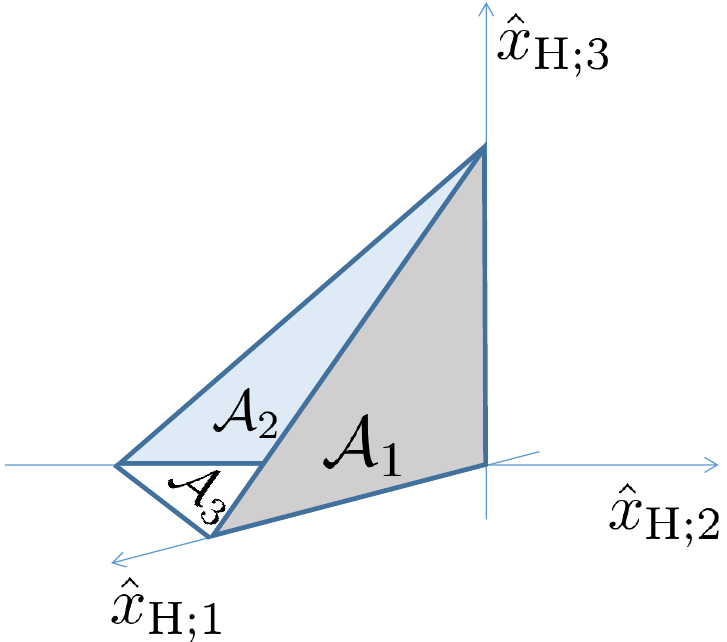}
\caption{A typical filling in {\bf Step 2$_{q=2}$}. We fill the $D^3$ bounded by three $D^2$ with labels $\mathcal{A}_1=\{\hat{x}_\text{H;1},\hat{x}_\text{H;3}\}$, $\mathcal{A}_2=\{-\hat{x}_\text{H;2},\hat{x}_\text{H;3}\}$ and $\mathcal{A}_3=\{\hat{x}_\text{H;1},-\hat{x}_\text{H;2}\}$ and these labels obviously satisfy the requirement dim[Span$(\mathcal{A}_i)\,\cap\,$Span$(\mathcal{A}_j)]=q-1=1$ and $|\mathcal{A}_i\cap\mathcal{A}_j|=q-1=1$ for any $i\neq j$.
We assign a new label to this newly filled $D^3$ by $\mathcal{A}=\mathcal{A}_1\cup\mathcal{A}_2=\{\hat{x}_\text{H;1},-\hat{x}_\text{H;2},\hat{x}_\text{H;3}\}$.}
\label{filling}
\end{figure}
The generalization to arbitrary $n$ is straightforward.
In {\bf Step 1} we introduce a $(d+n)$-dimensional invertible system by putting $A_d$ and $\bar{A}_d$, respectively, on the two sublattices of a $n$-dimensional hypercubic lattice spanned by $\hat{x}_{H;1},\ldots,\hat{x}_{H;n}$.
We also prepare an invertible system in the trivial phase $[E_{d+n}]$ by putting $E_d$ on the $n$-dimensional hypercubic lattice.

In {\bf Step 2$_0$},
we connect the two systems introduced in {\bf Step 1} by $2n$ distinct ways, each of which corresponds to taking one of the $2n$ ``dimerization'' directions
$\pm\hat{x}_{H,i}$, $i=1,\ldots,n$.
In {\bf Step 2$_1$} we fill in the $2n(n-1)$ loop holes bounded by $S^1$ formed by two paths whose dimerization directions $\vec{v}$ and $\vec{w}$ are not parallel ($\vec{v},\vec{w}\in\{\pm\hat{x}_{H;1},\ldots,\pm\hat{x}_{H;n}\}$),
using the contraction procedure as in FIG.~\ref{sphere}.
Then we label this filled area (a 2-disk) $D^2$ by a set 
$\mathcal{A}=\{\vec{v},\vec{w}\}$ and the vector space $\mbox{Span}(\mathcal{A})=\{c_1\vec{v}+c_2\vec{w} \,|\, c_1, c_2\in\mathbb{R}\}$.
In {\bf Step 2$_q$} ($q=2,\ldots,{n-1}$),
we fill in any one ``hole'' $D^{q+1}$ bounded by $(q+1)$ of the filled $D^{q}$'s constructed in {\bf Step 2$_{q-1}$}.
For any pair of these selected $D^{q}$'s labels,
say $\mathcal{B}$ and $\mathcal{C}$,
their spanned vector spaces satisfy the following two conditions: $\text{dim}[\text{Span}(\mathcal{B})\,\cap\,\text{Span}(\mathcal{C})]=q-1$ and $|\mathcal{B}\,\cap\,\mathcal{C}|=q-1$; see FIG.~\ref{filling} for $q=2$.
The filling uses a higher-dimensional generalization of the loop contraction procedure in FIG.~\ref{sphere}.
Then we label this filled area $D^{q+1}$ by the union $\mathcal{B}\cup\mathcal{C}$.
Repeat inductively until {\bf Step 2$_{{n-1}}$}.
This procedure gives $J_A(\vec{\tau})\in[S^{n},\text{\p{d+n}}]_0$, where $\vec{\tau}$ parametrizes $S^n$.
{\bf Step 3} can be generalized without substantial changes from the $n=2$ case.
Namely, we observe that an $\vec{x}_H$-dependent gapped Hamiltonian has the same anomaly as $A_d$, and the group structure is preserved.

\begin{figure}[t]
\centering
\includegraphics[width=8.5cm,pagebox=cropbox,clip]{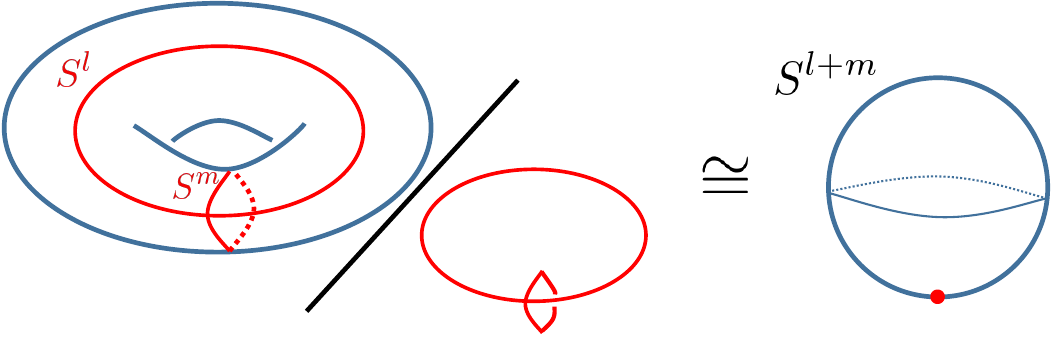}
\caption{The quotient $S^{l}\times S^m/(S^{l}\cup S^m)=S^{l+m}$ is done by collapsing to the south pole a certain pair of $S^{l}$ and $S^m$ which are mapped by $J_a(\vec{\tau},\vec{t})$ to $E_m$.  }
\label{wedge_product}
\end{figure}

It is in order here to discuss the relation between our isotropic approach and the suspension approach in~\cite{Wen:2021aa}
which constructs an element of $[S^{n},\text{\p{d}}]_0$ from an element $\varphi$ in $[S^{n-1},\text{\p{d-1}}]_0$.
The basic idea in the suspension is from the observation that the $S^n$ can be constructed from $S^{n-1}\times S^1$ as in FIG.~\ref{wedge_product} in which $l\equiv n-1$ and $m\equiv 1$.
The essential additional data needed in this construction is the trivializing procedure of $\varphi$ and its $(n-1)$-phase inverse $\bar{\varphi}$ to the trivial $(n-1)$-phase.
In our approach,
we make a direct connection to $[S^{0},\text{\p{d}}]_0$ from $[S^{n},\text{\p{d+n}}]_0$,
but in the above inductive procedure,
we need to know all $(q-1)$-phase trivializations involved in the above {\bf Step $2_q$} with $q=1,2,\cdots,{n-1}$.
Therefore,
the total numbers of trivializations to know in advance match between these two approaches.

We now show that Eq.~(\ref{pik}) can be further refined:
\begin{eqnarray}\label{pi}
[S^{n},\text{\p{d+n}}]_0=[S^{0},\text{\p{d}}]_0.
\end{eqnarray}
This equation implies that our construction is complete and surjective without losing any generality.
We can rephrase Eq.\ (\ref{pi}) as a dimension ladder:
\begin{equation}\label{ladder}
[S^{n},\text{\p{d+n}}]_0=[S^{n-1},\text{\p{d+n-1}}]_0.
\end{equation}

Our derivation of Eq.\ (\ref{pi}) goes as follows.
Let us take any element in {$[S^{n},\text{\p{d+n}}]_0$, which is represented by} a mapping ${J}: S^n\rightarrow\mathcal{P}(\text{Inv}_{d+n}^G)$ with a certain point $\vec{\tau}_0$ (e.g., taken as the south pole) on $S^n$ mapped to some atomic $E_{d+n}=\mathcal{H}[I_{d+n}^\text{tr}]$.
We can take $E_{d+n}$ as conventional co-$(d+n)$ atomic by the deformation Eq.~(\ref{unity}).
We construct its corresponding modulating Hamiltonian $\tilde{H}_J(\vec{x}_H)=\mathcal{H}[{J}(\vec{x}_H/L_H)]$, where $\vec{x}_H$ is a vector on $S^n$ with radius $L_H$.
We first adiabatically squeeze $\tilde{H}_J(\vec{x}_H)$ such that it is $E_{d+n}$ for most $\vec{x}_H$ except in a small area (e.g., around the north pole) of $S^n$ of size $(\lambda\xi)^n$, where $\xi$ is a finite correlation length and the parameter $\lambda$ is chosen as $\lambda\gg1$ to ensure a nonzero gap.
Therefore,
we can choose a sufficiently large $L_H\gg\lambda\xi$ to keep the modulating Hamiltonian to be gapped with a unique ground state.
Suppose that this squeezed $\tilde{H}_J(\vec{x}_H)$ have a $(d+n-1)$-dimensional boundary.
The anomaly from its boundary modes cannot match the boundary anomaly of any $m$-dimensional invertible phase if $m>d$,
because we have squeezed the variations along $n$ dimensions of $S^n$.
Thus its boundary anomaly must correspond to some bulk invertible phase $[A_d]$ in $[S^{0},\text{\p{d}}]_0$,
which is a correspondence from the left-hand side of Eq.~(\ref{pi}) to its right-hand side.

Finally,
we argue that,
if any two mappings $J_A$ and $J_{A'}$ from $S^n\rightarrow\mbox{\p{d+n}}$ give, under the above correspondence, $[A_d]$ and $[A_d']$, respectively,
and if $[A_d]=[A_d']$,
then they are in the same $n$-phase.
Since the above squeezing procedure is an adiabatic process,
$\sum_{\vec{x}_H}\tilde{H}_{J_A}(\vec{x}_H)$
and $\sum_{\vec{x}_H}\tilde{H}_{J_{A'}}(\vec{x}_H)$
are both in the same phase,
which is $[A_d]$.
Thus,
\begin{eqnarray}
\left(\sum_{\vec{x}_H}\tilde{H}_{J_A}(\vec{x}_H)\right)\h E_{d+n}
\Leftrightarrow
\left(\sum_{\vec{x}_H}\tilde{H}_{J_{A'}}(\vec{x}_H)\right)\h E_{d+n}'
\label{J_A = J_A'}\nonumber
\end{eqnarray}
through a path
$\tilde{F}_{s\in[0,1]}\in\mbox{\p{d+n}}$.
We \textit{assume} that this path can be chosen such that each point (that is a modulating Hamiltonian) on it can be continuously mapped back through:
\begin{eqnarray}\label{backward}
F(s,\vec{\tau})\equiv \mathcal{H}^{-1}(\tilde{F}_{s}(\vec{\tau}L_H)),
\end{eqnarray}
which defines $F:[0,1]\times S^n\rightarrow\mbox{\p{d+n}}$ that realizes the connection between $J_A+J_E$ and $J_{A'}+J_{E'}$; they are in the same $n$-phase.
Together with the result in the last paragraph,
we come to the conclusion of Eq.~(\ref{pi}).

However,
we should keep in mind that the assumed path $\tilde{F}_s$ realizing Eq.~(\ref{J_A = J_A'})
can be highly model-dependent in general,
so that the loop contraction at the upper right of FIG.~\ref{sphere} may depend on the details of the lattice model as well.
Moreover, a Hamiltonian on the contraction path may have coupling constants that rapidly vary in space on the scale of lattice constants; such a Hamiltonian will not have a field-theory description.
Nevertheless, for one of the simplest nontrivial case where $d=0$ and $n=2$,
we will show in the coming Sec.~\ref{sphere_zero_dim} that the elements in $[S^2,\text{\p{2}}]_0$ can be explicitly constructed in a model-independent way.

We have constrained $\vec{\tau}_0$ to be mapped to the trivial phase on the left-hand side of Eq.~(\ref{pi}).
When we relax this constraint, we have the extra freedom of choosing different phases to which $\vec{\tau}_0$ is mapped, and
this choice can be precisely accounted for by $[S^0,\text{\p{d+n}}]_0$, whose elements are all the $(d+n)$-dimensional invertible phases.
Let us consider an arbitrary element $\Phi:S^n\rightarrow\mathcal{P}(\text{Inv}_{d+n}^G)$ in $[S^n,\mathcal{P}(\text{Inv}_{d+n}^G)]$ with $\Phi(\vec{\tau}_0)=A_{d+n}$.
Then $(J_{\bar{A}}+\Phi)(\vec{\tau}_0)=E_{d+n}$, where $J_{\bar{A}}$ denotes the constant map that maps the whole $S^n$ to $\bar{A}_{d+n}$ and ``+'' is the group addition defined by pointwise stacking of Hamiltonian as before.
Thus, 
such a mapping $(J_{\bar{\cdot}}+\cdot)$ maps $[S^n,\mathcal{P}(\text{Inv}_{d+n}^G)]$ back to $[S^n,\text{\p{d+n}}]_0$,
and therefore $[S^n,\mathcal{P}(\text{Inv}_{d+n}^G)]$ is exactly $[S^0,\text{\p{d+n}}]_0$ copies of $[S^n,\text{\p{d+n}}]_0=[S^0,\text{\p{d}}]_0$, as reflected in the following direct product of groups
\begin{eqnarray}\label{pifinal}
[S^n,\mathcal{P}(\text{Inv}_{d+n}^G)] = [S^0,\text{\p{d}}]_0\!\!\times\!\![S^0,\text{\p{d+n}}]_0.
\end{eqnarray}
We note, however, that this equation cannot have a ladder relation as Eq.~(\ref{ladder}).

So far we have considered $[S^n,\mathcal{P}(\text{Inv}_{m}^G)]_0$ only when $m\geq n$.
In fact, the remaining case $l\equiv n-m>0$ is also accessible by our construction, yielding the following result:
\begin{eqnarray}\label{pi_berry}
[S^{l+m},\mathcal{P}(\text{Inv}_{m}^G)]_0=[S^l,\mathcal{P}(\text{Inv}_{0}^G)]_0
\end{eqnarray}
to be clarified below.
Let us take an element of the right-hand side,
which is a mapping $\mathcal{H}_a(\vec{\tau}):\vec{\tau}\in S^{l}\rightarrow\text{\p{0}}$.
With its inverse ${\mathcal{H}}_{\bar{a}}$, the combined mapping $\mathcal{H}_a(\vec{\tau})+{\mathcal{H}}_{\bar{a}}(\vec{\tau})$ is
in the same $l$-phase as $J_E$, which is a constant mapping from $S^l$ to a co-1 atomic Hamiltonian $E_0$.
For each \textit{fixed} ${\vec{\tau}}\in S^{l}$,
we define a Hamiltonian $H_{a\bar{a}}(\vec{\tau})\in\mbox{\p{m}}$
by putting $\mathcal{H}_a(\vec{\tau})$ and $\bar{\mathcal{H}}_{\bar{a}}(\vec{\tau})$, respectively, on the two sublattices of $m$-dimensional hypercubic lattice, in analogy with the construction discussed above [see FIG.~\ref{loop_mod} for $m=1$ and FIG.~\ref{sphere} for $m=2$, where $A_d$ and $\bar{A}_d$ are used instead of $\mathcal{H}_a(\vec{\tau})$ and $\bar{\mathcal{H}}_{\bar{a}}(\vec{\tau})$].
Varying $\vec{\tau}\in S^{l}$,
this gives a mapping $H_{a\bar{a}}:S^{l}\rightarrow\text{\p{m}}$.
We place $H_{a\bar{a}}(\vec{\tau})$ at the north pole of $S^m$.
For a fixed $\vec{\tau}$, we can dimerize the pairs in $H_{a\bar{a}}(\vec{\tau})$ in $2m$ directions on $S^m$ and fill the holes (as in FIG.~\ref{sphere} when $m=2$),
obtaining a mapping from $S^m$ to \p{m} for fixed $\vec{\tau}$; we parametrize this $S^m$ by $\vec{t}$.
With varying $\vec{\tau}$,
we have a mapping $J_a(\vec{\tau},\vec{t})$ where $(\vec{\tau},\vec{t})\in S^{l}\times S^{m}$.
We note that \{{the} south pole of $S^l$\}$\times S^m$ and $S^l\times$\{{the} south pole of $S^m$\} are mapped {by $J_a$} to the same single point $E_m$ {in \p{m}}.
It means that we can actually quotient them out $S^{l}\times S^{m}/(S^{l}\cup S^m)\cong S^{l+m}$, as in FIG.~\ref{wedge_product}, which induces the resultant mapping $\bar{J}_a:S^{l+m}\rightarrow\text{\p{m}}$.
It is the desired element of the left-hand side of Eq.~(\ref{pi_berry}), {in which $m$ is any positive integer.}

\section{Lattice Constructions of $[S^2,\text{\p{2}}]_0$}\label{sphere_zero_dim}


\begin{figure}[h]
\centering
\includegraphics[width=8.8cm,pagebox=cropbox,clip]{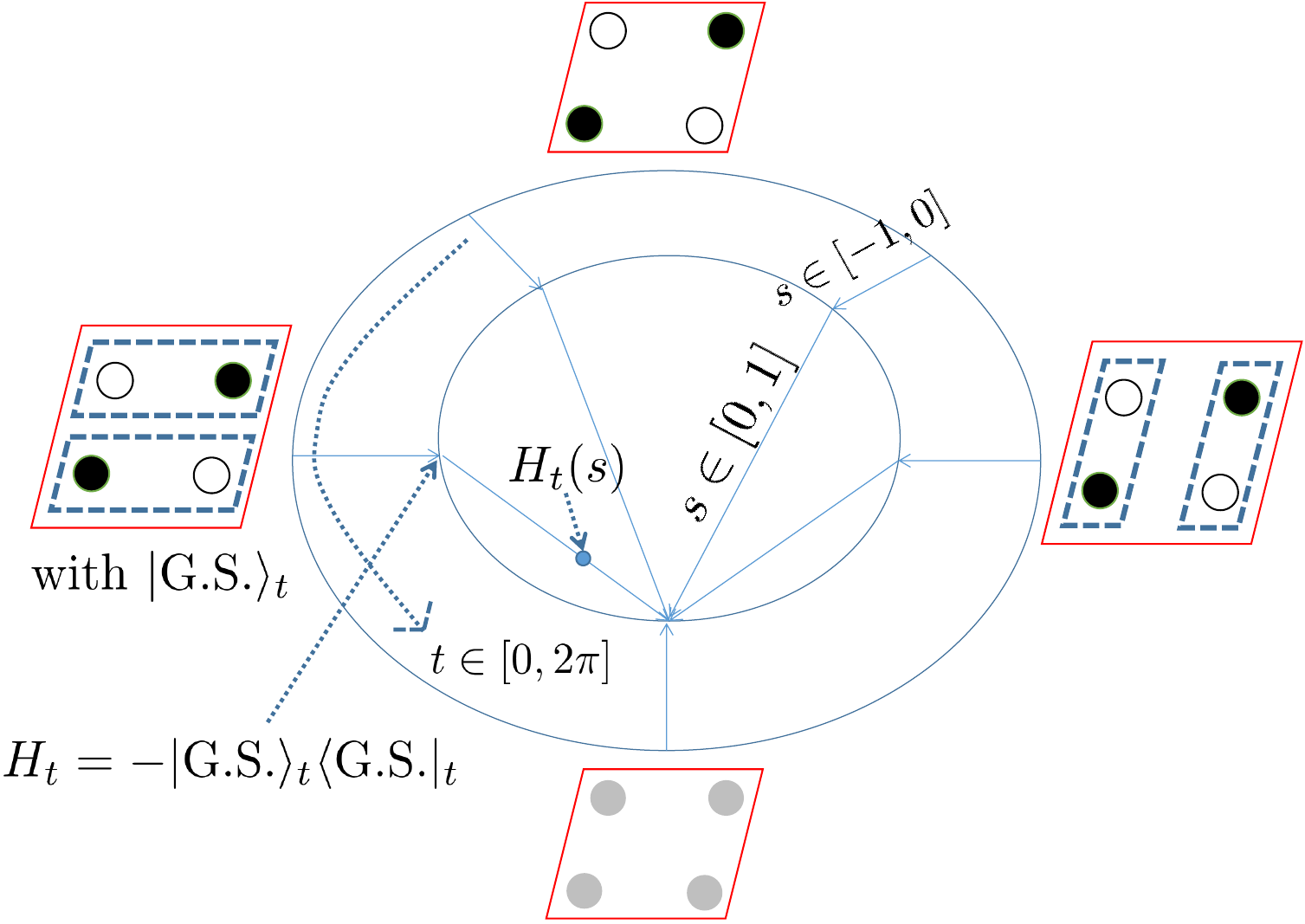}
\caption{To fill each loop hole in FIG.~\ref{sphere}~(left),
we only need to consider the contraction of loops of square Hamiltonians.
Here $|\text{G.S.}\rangle_t$ is the ground state of the Hamiltonian on the outer circle.}
\label{sphere_1}
\end{figure}

In this section we will explicitly construct the elements of $[S^2,\text{\p{2}}]_0$,
i.e., $[S^2,\text{\p{2+d}}]_0$ with $d=0$.
We start with FIG.~\ref{sphere} with a zero-dimensional system $A_0$ depicted as solid dots, its inverse $\bar{A}_0$ open dots, and $E_0$ grey dots.
A closer look at the left panel of FIG.~\ref{sphere} shows that every point on the loop bounding one of the four holes of $S^2$ is composed of decoupled squares as shown in FIG.~\ref{sphere}~(left).
Hence,
in order to fill in these holes,
we only need to consider contractions of Hamiltonian loops within each separate square.

This contraction procedure is schematically shown in FIG.~\ref{sphere_1}.
The loop of Hamiltonian for a square is parametrized by $t\in[0,2\pi]$, where the points $t=0$ and $t=2\pi$ correspond to the north pole (decoupled $A_0$ and $\bar{A}_0$) and the point $t=\pi$ corresponds to the south pole (a square Hamiltonian consisting of four $E_0$'s).

In the first step, the outer loop in FIG.~\ref{sphere_1} is deformed to the inner loop by adiabatically changing, for each $t$, the ground-state energy to $-1$ and the energy of all the excited states to zero.
This adiabatic deformation parametrized by $s\in[-1,0]$ from the outer loop ($s=-1$) to the inner loop ($s=0$) can be done for any zero-dimensional system using the one-parameter Hamiltonian:
\begin{eqnarray}\label{moving}
H(s)&=&\left[(-s)E_\text{G.S.}-(s+1)\right]|\text{G.S.}\rangle\langle\text{G.S.}|
\nonumber\\
&&+\sum_{i=1}[(-s)E_i]|\Psi_i\rangle\langle\Psi_i|,
\end{eqnarray}
where $|\text{G.S.}\rangle$ is the ground state with energy $E_\text{G.S.}$,
$|\Psi_i\rangle_{i=1,2,\cdots}$ are excited states with energy $E_i$, and
we have suppressed the parameter $t$ for brevity.
Note that the Hamiltonian at $s=0$ is simply
$H(0)=-|\text{G.S.}\rangle \langle\text{G.S.}|$.

In the next step,
the inner loop ($0\le t\le2\pi$) is contracted to the point ($t=\pi$) using the Hamiltonian with the parameter $s\in[0,1]$:
\begin{eqnarray}\label{in_out}
H_t(s)&\equiv&-\frac{1}{\mathcal{N}_{t,s}}
\left[\cos\left(\frac{\pi s}{2}\right)|\text{G.S.}\rangle_t+\sin\left(\frac{\pi s}{2}\right)|\text{G.S.}\rangle_\pi\right]\nonumber\\
&&\times\left[\cos\left(\frac{\pi s}{2}\right)\langle\text{G.S.}|_t+\sin\left(\frac{\pi s}{2}\right)\langle\text{G.S.}|_\pi\right],
\end{eqnarray}
where the normalization constant ($\mathcal{N}_{t,s}>0$), defined by
\begin{equation}
\mathcal{N}_{t,s}\equiv 1+\text{Re}\left({}_\pi\langle\text{G.S.}|\text{G.S.}\rangle_t\right)\sin(\pi s),
\end{equation}
keeps the ground-state energy to be $-1$.

It should be noted that in the above construction we have assumed a continuously parametrized ground states $|\text{G.S.}\rangle_{t\in[0,2\pi]}$,
which necessarily requires $|\text{G.S.}\rangle_0=|\text{G.S.}\rangle_{2\pi}$.
In fact,
this assumption can always be satisfied;
for example,
if $|\text{G.S.}\rangle_0=\exp(i\alpha)|\text{G.S.}\rangle_{2\pi}$,
then we can choose a new $|\text{G.S.}\rangle_t$ to be $\exp(-i\alpha t/2\pi)|\text{G.S.}\rangle_t$, so that the discontinuity is removed.
Furthermore,
two neighboring loops may have $|\text{G.S.}\rangle_t$'s that differ by a phase factor on the meridian shared by the two loops,
since we only require their parent Hamiltonians to be the same.

Unfortunately, the above argument cannot be directly generalized to
$[S^2,\text{\p{2+d}}]_0$ with $d>0$, since the adiabatic deformations in Eqs.~(\ref{moving}) and (\ref{in_out}) generically break the locality of higher-dimensional Hamiltonian.
In fact, the contraction procedure appears to be model- and symmetry-dependent already at $d=1$ as constructed in~\cite{Yao:2022vh}.
We also note that the isotropic constructions of $[S^n,\text{\p{n}}]_0$ for $n>2$ can be of future interest.

\section{Crystalline topological phases by generalized magnetic translations}\label{crystalline}
In this section,
we will give several applications of the modulating Hamiltonian approach to possible crystalline topological phases protected by a generalized magnetic translation,
which is a combination of the lattice translation and a symmetry that can transform a topological phase to its inverse.

We will first give a concrete example of a two-dimensional quantum Hall system as motivation.
Later,
we will see that the Lieb-Schultz-Mattis theorem and its magnetic translation version can be obtained as special cases of our general result. 

\subsection{Motivating applications}\label{sec: applicationsA}
Let us assume $G_\text{onsite}=\mathrm{U}(1)$ and $d=2$.
We consider the two-dimensional electronic system with
\begin{eqnarray}
A_{d;G_\text{onsite}}&=&\{\sigma_\text{H}=1\};\nonumber\\
\bar{A}_{d;G_\text{onsite}}&=&\mathcal{T}(A_{d;G_\text{onsite}})=\{\sigma_\text{H}=-1\},
\end{eqnarray}
where $\{\sigma_\text{H}=1\}$ is a Chern insulator with the electric Hall conductance $\sigma_\text{H}=1$ in units of $e^2/h$, $\bar{A}_{d;G_\text{onsite}}$ is obtained by the time-reversal transformation $\mathcal{T}$ of $A_{d;G_\text{onsite}}$, which can be  realized, for instance, in the Haldane's honeycomb model~\cite{Haldane:1988uf}.
Later,
we will also take into consideration the integer thermal quantum Hall conductance for the sake of completeness.


We claim that the following staggering (decoupled) stacking
\begin{eqnarray}\label{cSPT}
&&\mathcal{H}_{d+1}\nonumber\\&&\equiv\cdots \h A_{d;G_\text{onsite}} \h \bar{A}_{d;G_\text{onsite}} \h A_{d;G_\text{onsite}} \h \bar{A}_{d;G_\text{onsite}} \h\cdots\nonumber\\
\end{eqnarray}
is a nontrivial topological phase protected by the symmetry:
\begin{eqnarray}
G=\{G_\text{onsite},M_\mathcal{T}\},
\end{eqnarray}
where $M_\mathcal{T}$ is the so-called magnetic translation
\begin{eqnarray}
M_\mathcal{T}\equiv\text{Tr}\circ\mathcal{T},
\end{eqnarray}
defined as the composition of the translation symmetry ``Tr'' along the stacking direction in Eq.~(\ref{cSPT}) and the time reversal $\mathcal{T}$.
(We have used $\mathcal{H}_{d+1}$ for lattice Hamiltonian in Eq.~(\ref{cSPT}),
which has no relation with the Hamiltonian mapping $\mathcal{H}$ before.)

\begin{figure}[h]
\centering
\includegraphics[width=8.5cm,pagebox=cropbox,clip]{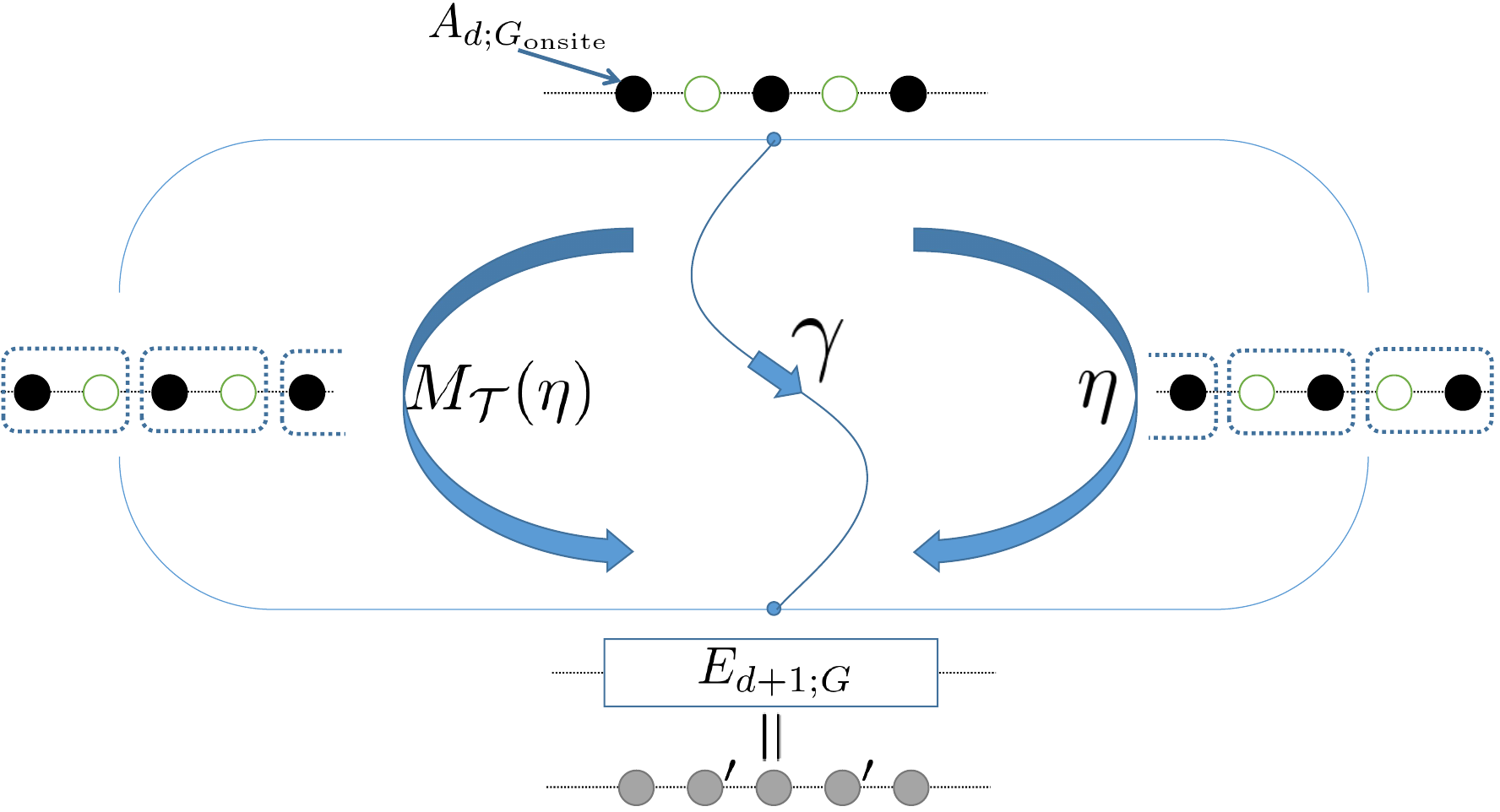}
\caption{The illustration of various paths.}
\label{app}
\end{figure}

We prove the above claim by contradiction. Let us assume that $\mathcal{H}_{d+1}$ is in a $G$-trivial phase.
Then there is a $G$-symmetric, thus $G_\text{onsite}$-symmetric, path from $\mathcal{H}_{d+1}$ to a $G$-trivial phase $E_{d+1;G}$.
This $E_{d+1;G}$ is necessarily an atomic insulator; we can choose $E_{d+1;G}$ to have the form
\begin{eqnarray}
&&E_{d+1;G}\nonumber\\
&&=\cdots \h E_{d;G_\text{onsite}} \h E'_{d;G_\text{onsite}} \h E_{d;G_\text{onsite}} \h E'_{d;G_\text{onsite}} \h \cdots,\nonumber
\end{eqnarray}
with some $G_\text{onsite}$-trivial Hamiltonian satisfying $E'_{d;G_\text{onsite}}\equiv\mathcal{T}E_{d;G_\text{onsite}}\mathcal{T}^{-1}$ due to the $M_\mathcal{T}$ symmetry.
The $G$-symmetric path from $\mathcal{H}_{d+1}$ to $E_{d+1;G}$ is denoted as $\gamma$.
We also consider a $G_\text{onsite}$-symmetric path involving pairwise trivialization from $\mathcal{H}_{d+1}$ to $E_{d+1;G}$, which is denoted by $\eta$; see FIG.~\ref{app}.
By acting on $\eta$ with $M_{\mathcal{T}}$,
we obtain another path $M_{\mathcal{T}}(\eta)$ from $\mathcal{H}_{d+1}$ to $E_{d+1;G}$.
In other words, we define the $\mathcal{T}$-symmetry operation on a path by $M_\mathcal{T}(\eta)(\tau)\equiv M_\mathcal{T}^{-1}[\eta(\tau)]M_\mathcal{T}$ to produce the new path $M_\mathcal{T}(\eta)$.
The above setting of paths is shown in FIG.~\ref{app},
in which the grey dots (with prime) represent $E_{d;G_\text{onsite}}$ (with prime).
For these paths, we construct their modulating Hamiltonians as in Eq.\ (\ref{modloop}).

We denote the composition (or connection) of paths $a$ and $b$ by $a*b$, which reads ``first path $a$ and then path $b$'':
\begin{eqnarray}
a*b(\tau)=\left\{\begin{array}{cc}a(2\tau),&\tau\in[0,1/2];\\b(2\tau-1),&\tau\in(1/2,1].\end{array}\right.
\end{eqnarray}

\begin{figure}[h]
\centering
\includegraphics[width=8.5cm,pagebox=cropbox,clip]{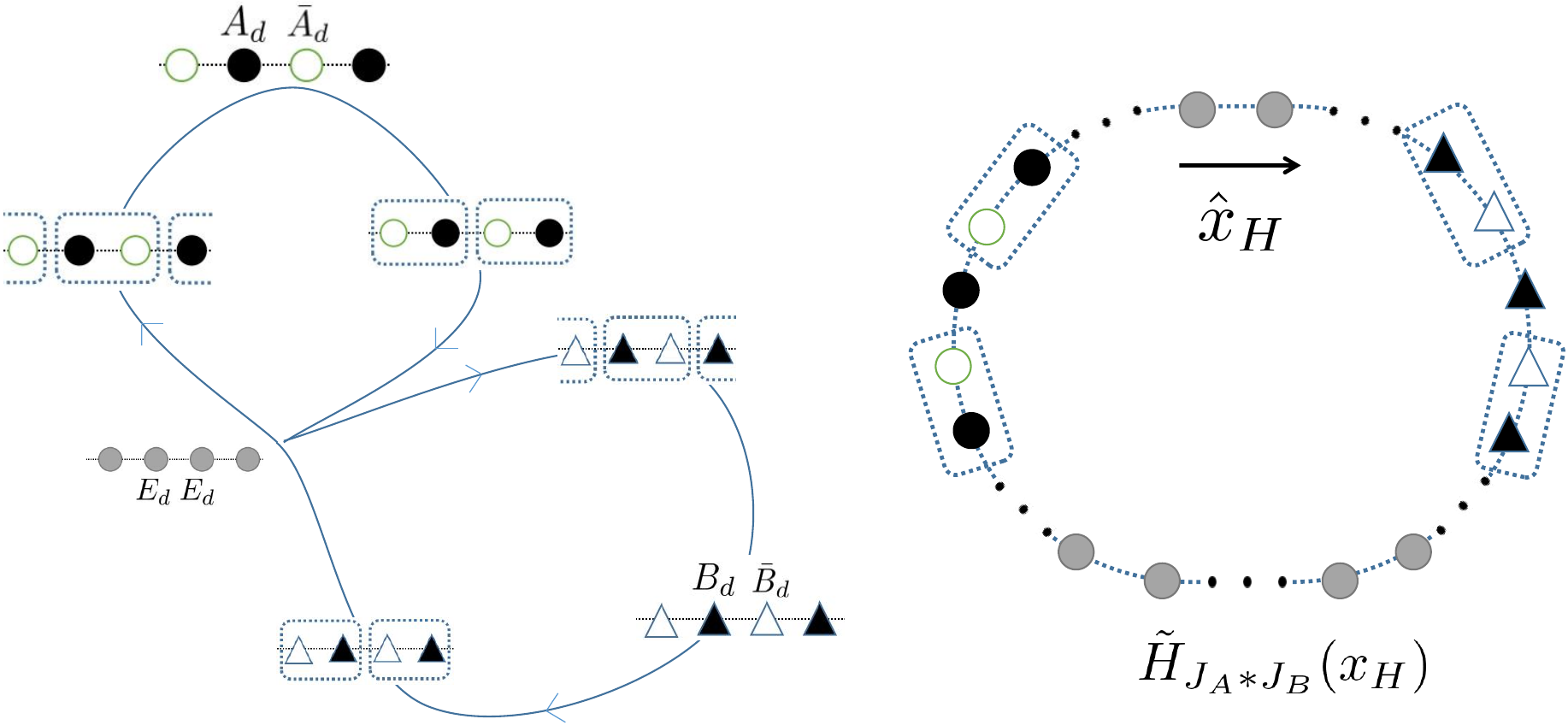}
\caption{The modulating Hamiltonian $\tilde{H}_{J_A*J_B}$ (right) of the composite loop  $J_A*J_B$ (left) has the same topological invariant as $\tilde{H}_{J_A}\h\tilde{H}_{J_B}$ after the modulating direction is compactified.}
\label{loop_star}
\end{figure}
\begin{figure}[h]
\centering
\includegraphics[width=8.5cm,pagebox=cropbox,clip]{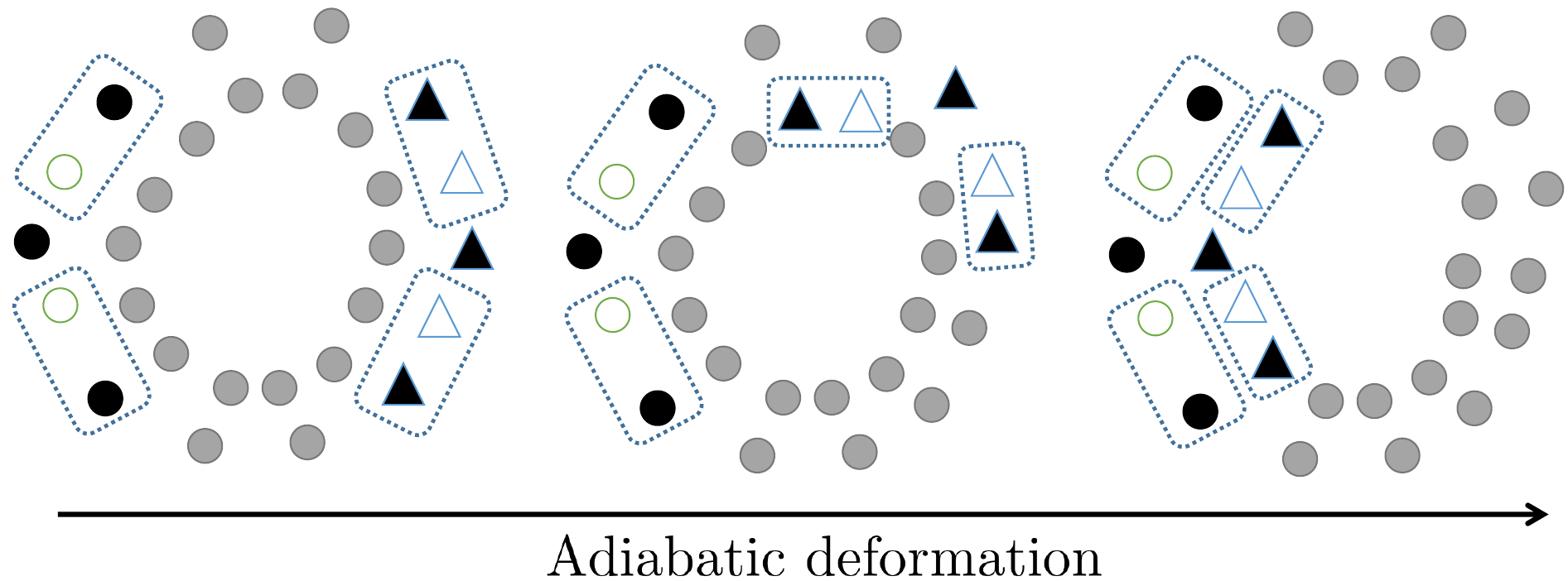}
\caption{Illustration of the adiabatic deformation of $[\tilde{H}_{J_A*J_B}]=[\tilde{H}_{J_A*J_B+J_E}]$ to $[\tilde{H}_{J_A}\text{-}\tilde{H}_{J_B}]$.}
\label{loop_deformation}
\end{figure}
By the definition of $a*b$,
the modulating Hamiltonian {$\tilde{H}_{J_A*J_B}$ associated with $J_A*J_B$ is in the same $0$-phase as $\tilde{H}_{J_A}\h\tilde{H}_{J_B}$ after the modulating direction is compactified (Fig.~\ref{loop_star}):
\begin{eqnarray}\label{loop_add}
[\tilde{H}_{J_A*J_B}]=[\tilde{H}_{J_A}\h\tilde{H}_{J_B}].
\end{eqnarray} }\noindent
{This implies that {the fundamental group has the same abelian group structure as the stacking of Hamiltonians, as we explain below with Fig.~\ref{loop_deformation} showing $[\tilde{H}_{J_A*J_B}]=[\tilde{H}_{J_A}\text{-}\tilde{H}_{J_B}]$.
Since the equivalence relation is defined up to the addtion of a trivial phase} in Eq.~(\ref{1_phase}),
we can stack a {closed loop} of (decoupled) $E_d$'s to the $\tilde{H}_{J_A*J_B}$.
The top and the bottom part of this {modulating} Hamiltonian are now decoupled $E_d$'s {in Fig.~\ref{loop_deformation},} and we imagine the triangles slowly ``move'' to the gray dots right {below} the black/white dots.
Such a {``move''} is realized by adiabatic deformation of {$\tilde{H}_{J_A*J_B}$,
which becomes eventually $\tilde{H}_{J_A}\h\tilde{H}_{J_B}$.}
The above argument {can also be generalized} to higher homotopies.
It should be noted that {$J_A$ and $J_B$} above must be loops rather than arbitrary paths.} 

For the sake of clearness,
we will suppress the tilde over the modulating Hamiltonians in the following discussions.
We also denote the inverse of a path $\alpha$ by $\bar{\alpha}:\bar{\alpha}(\tau)\equiv\alpha(1-\tau)$ with $\tau\in[0,1]$.
Since $a*\bar{a}$ is deformable to the constant map,
we can conclude that $H_{a}\h H_{\bar{a}}$ and $E_{d;G_\text{onsite}}$ are in the same $0$-phase for any loop $a$ or
\begin{eqnarray}
[H_{\bar{a}}]=[\bar{H}_a],
\label{[H_bar]=[bar H]}
\end{eqnarray}
where the bar over the Hamiltonian denotes its inverse in the classification of the topological phases.
From the previous consideration below Eq.\ (\ref{modloop}),
we obtain
\begin{equation}
[H_{M_{\mathcal{T}}(\eta)*\bar{\eta}}]=[ A_{d;G_\text{onsite}}].
\label{[A_onsite]}
\end{equation}
We see from FIG.~\ref{app} that the connected path $M_\mathcal{T}(\eta)*\bar{\gamma}$ is transformed by the $\mathcal{T}$-operation to $\eta*\bar{\gamma}$, i.e.,
\begin{eqnarray}
M_{\mathcal{T}}H_{M_{\mathcal{T}}(\eta)*\bar{\gamma}}M_{\mathcal{T}}^{-1}=H_{\eta*\bar{\gamma}},
\end{eqnarray}
where we have used that fact that $\gamma$ is $M_{\mathcal{T}}$-invariant.
Moreover, the composition of $M_\mathcal{T}(\eta)*\bar{\gamma}$ and the inverse path of $\eta*\bar{\gamma}$ is equivalent to the connected path $M_\mathcal{T}(\eta)*\bar{\eta}$,
\begin{eqnarray}
M_{\mathcal{T}}(\eta)*\bar{\gamma}*\overline{\eta*\bar{\gamma}}=M_{\mathcal{T}}(\eta)*\bar{\eta},
\label{connected path}
\end{eqnarray}
since $\bar{\gamma}*\gamma$ is deformable to the identity map.
It follows from Eqs.~(\ref{loop_add}),~
(\ref{[H_bar]=[bar H]}),~(\ref{[A_onsite]})~and~(\ref{connected path}) that
\begin{eqnarray}
[A_{d;G_\text{onsite}}]&=&[H_{M_{\mathcal{T}}(\eta)*\bar{\gamma}}\h{H}_{\overline{\eta*\bar{\gamma}}}]\nonumber\\
&=&[H_{M_{\mathcal{T}}(\eta)*\bar{\gamma}}\h\bar{H}_{\eta*\bar{\gamma}}]\nonumber\\
&=&[H_{M_{\mathcal{T}}(\eta)*\bar{\gamma}} \h \overline{M_{\mathcal{T}}H_{M_{\mathcal{T}}(\eta)*\bar{\gamma}}M_{\mathcal{T}}^{-1}}].
\end{eqnarray}
Since $M_\mathcal{T}$ transformation flips the sign of $\sigma_\text{H}$ of $H_{M_{\mathcal{T}}(\eta)*\bar{\gamma}}$ and the bar operation inverts its $\sigma_\text{H}$ back,
we have
\begin{eqnarray}
[H_{M_{\mathcal{T}}(\eta)*\bar{\gamma}}]=[\sigma_\text{H}=1/2],
\end{eqnarray}
which is impossible for an invertible Hamiltonian~\cite{Laughlin:1981aa}.
Hence, we conclude that $\mathcal{H}_{d+1}$ in Eq.~(\ref{cSPT}) is in a nontrivial topological phase protected by $G=\{G_\text{onsite},M_\mathcal{T}\}$.

\begin{figure}[t]
\centering
\includegraphics[width=8.8cm,pagebox=cropbox,clip]{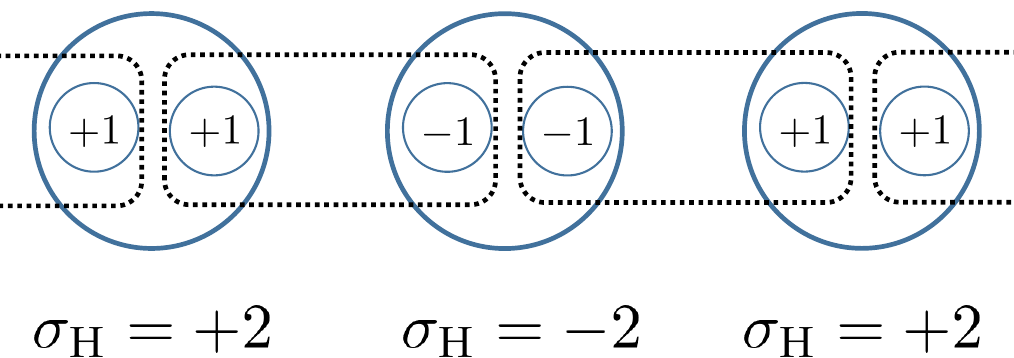}
\caption{A $G$-symmetric trivialization in the case $A_{d;G_\text{onsite}}$ is divisible by $2$: we have suppressed $\kappa_\text{H}$'s and the dashed triangles means a trivialization of a pair of $\sigma_\text{H}=-1$ and $\sigma_\text{H}=+1$.}
\label{neutral}
\end{figure}

On the other hand, if $A_{d;G_\text{onsite}}=\{\sigma_\text{H}=2\}$, $\mathcal{H}_{d+1}$ in Eq.\ (\ref{cSPT}) is in a trivial phase,
and there is an adiabatic path to $E_{d+1}$.
This path can be easily constructed by splitting $\{\sigma_\text{H}=2\}$ into two $\{\sigma_\text{H}=1\}:=A'_{d}$ and also its inverse $\{\sigma_\text{H}=-2\}$ into two $\{\sigma_\text{H}=-1\}=\bar{A'_{d}}$,
\begin{equation}
\mathcal{H}_{d+1}=\cdots \h (A'_{d}\h A'_{d}) \h (\bar{A'_{d}}\h\bar{A'_{d}})
\h  (A'_{d}\h A'_{d})\h (\bar{A'_{d}}\h\bar{A'_{d}}) \h \cdots.
\end{equation}
Trivializing the neighboring $A'_d\h\bar{A'_d}$ and $\bar{A'_d}\h A'_d$ yields $E_{d+1;G}$, as shown in FIG.~\ref{neutral}.
This trivialization can also be systematically obtained through lattice homotopy approach~\cite{Po:2017aa,Shiozaki:2022wm,Else:2020aa}.
Therefore, the classification of the $G$-symmetric Hamiltonian (\ref{cSPT}) has a $\mathbb{Z}_2$ structure.

\subsection{Inclusion of thermal Hall conductance}
In general,
U$(1)$-symmetric invertible electronic systems are characterized by two integer-valued topological invariants under an appropriate unit system $(\sigma_\text{H};\kappa_\text{H})$ which are constrained by~\cite{Lu:2012ux,Wang:2015vv,Seiberg:2016aa}
\begin{eqnarray}\label{ele_therm}
\sigma_\text{H}=\kappa_\text{H}\mod8,
\end{eqnarray}
{where $\kappa_\text{H}
$ represents thermal Hall conductance.}
Therefore,
the above consideration implies that the {classification of invertible phases under the symmetry $G$ (\ref{cSPT})} is actually given by
\begin{subequations}
\begin{equation}
\mathbb{Z}_2\times\mathbb{Z}_2
\end{equation}
generated by
\begin{equation}
\{(\sigma_\text{H}=1,\kappa_\text{H}=1),(\sigma_\text{H}=0,\kappa_\text{H}=8)\}.
\label{two generators}
\end{equation}
\end{subequations}
{The $\mathbb{Z}_2$ group structures follow from the generalization of the above discussion that
any doubled system is divisible by $2$, thereby resulting in a trivial crystalline topological phase in the same way as in FIG.~\ref{neutral}.
Thus,
we only need to look for the systems that satisfy the constraint~(\ref{ele_therm}) and are indivisible by $2$,
and they are exactly the two generators of those $\mathbb{Z}_2$'s.}

For bosonic systems,
let us use the convention that the fundamental charged boson to have charge $2$.
Combining the result of integer bosonic quantum Hall states~\cite{Senthil:2013tp} and Eq.~(\ref{ele_therm}),
we obtain
\begin{eqnarray}
\sigma_\text{H}=0\mod8,
\qquad
\kappa_\text{H}=0\mod8.
\end{eqnarray}
Thus the classification of the $G$-symmetric crystalline topological phases in the form of Eq.~(\ref{cSPT}) is also
\begin{eqnarray}
&&\mathbb{Z}_2\times\mathbb{Z}_2\text{ generated by }\nonumber\\
&&\{(\sigma_\text{H}=8,\kappa_\text{H}=0),(\sigma_\text{H}=0,\kappa_\text{H}=8)\}.
\end{eqnarray}

\subsection{Another interesting example: magnetic LSM theorem}
We can consider the case with $d=1$, $G_\text{onsite}=\mathbb{Z}_2\times\mathbb{Z}_2\text{ or }\mathcal{T}$, and
\begin{eqnarray}
A_{d;G_\text{onsite}}=\{\text{spin-1 antiferromagnetic Heisenberg chain}\}.\nonumber
\end{eqnarray}
The above argument is also applicable to this case, since the classification of $G_\text{onsite}$-symmetric invertible phases in $d=1$ is $\mathbb{Z}_2$, whose generator is the Haldane phase; i.e., there is no invertible phase whose double is the Haldane phase.

One of the interesting consequences is its boundary LSM theorem --- reproducing the LSM theorem with magnetic translation~\cite{YaoLiHsiehOshikawa}.

We can also replace $\mathcal{T}$ by the identity operator, and then the traditional LSM theorem follows.

\subsection{A short summary and more examples}
So far,
in the $G_\text{onsite}=\mbox{U}(1)$ case,
we have made use of the following properties:
\begin{itemize}
\item $\mathcal{T}$ flips $\sigma_\text{H}$ of any invertible phase to its opposite;
\item any invertible Hamiltonian with $\sigma_\text{H}=1$ is not in the same phase as two invertible phases, i.e., $\sigma_\text{H}=1$ is not divisible by 2.
\end{itemize}
In addition, $\mathcal{T}$ can be replaced by a reflection symmetry.

For the spin chains with $G_\text{onsite}=\mathbb{Z}_2\times\mathbb{Z}_2\text{ or }\mathcal{T}$,
we have relied on the fact that 
\begin{itemize}
\item the invertible-phase classification is only $\mathbb{Z}_2$,
\end{itemize}
so the Haldane phase is also indivisible by 2, and $\mathcal{T}$ must also transform one phase to its inverse (which is itself).

In general, for
any symmetry transformation $K$ satisfying
\begin{eqnarray}\label{flip}
K A_{d;G_\text{onsite}} K^{-1}=\bar{A}_{d;G_\text{onsite}}
\end{eqnarray}
for all $A_{d;G_\text{onsite}}$, we can conclude that the $(d+1)$-dimensional Hamiltonian
\begin{equation}
{\mathcal{H}_{d+1}=}\cdots \h A_{d;G_\text{onsite}} \h \bar{A}_{d;G_\text{onsite}} \h A_{d;G_\text{onsite}} \h \bar{A}_{d;G_\text{onsite}} \h\cdots
\end{equation}
is in  a nontrivial $G$-topological phase with
\begin{eqnarray}
G=\{G_\text{onsite},\text{Tr}\circ K\},
\end{eqnarray}
as long as $A_{d;G_\text{onsite}}$ is indivisible by $2$.
Furthermore,
its boundary mode gives a corresponding generalized $G$-LSM theorem as long as the symmetry $K$ is not destroyed by such a boundary cutting.
We call such Tr$\circ K$ as a \textit{generalized magnetic translation}.


It should be noted that 
the condition of ``indivisible by 2'' is necessary, to prevent from the trivialization in FIG.~\ref{neutral}.

We list several more examples below:
\begin{itemize}
\item $G_\text{onsite}=\text{U}(1)$; $A_{0;G_\text{onsite}}=\{\text{charge}=1\}$; $K=$ Charge Conjugation: One-dimensional crystalline topological insulators;
\item $G_\text{onsite}=\text{U}(1)$; $A_{2;G_\text{onsite}}=\{\sigma_\text{H}=1,\kappa_\text{H}=1\}$; $K=$ Reflection with the reflection line within each $A_{2;\text{U}(1)}$ and $\bar{A}_{2;\text{U}(1)}$: Three-dimensional crystalline topological insulators; 
\item $G_\text{onsite}=1$; $A_{2;G_\text{onsite}}=\{p+ip$ superconductor\}~\cite{Read:2000vz}; $K=$ Time Reversal: Three-dimensional crystalline topological superconductors.
\end{itemize}

\section{Discussions}
The isomorphisms in Eq.~(\ref{ladder}) is a direct consequence of suspension isomorphism if we assume that \p{d} is homotopy equivalent to the loop space of \p{d+1},
i.e., \{\p{d}\}$_{d=0,1,\cdots}$ forming an $\Omega$-spectrum~\cite{Kitaev:2011aa}.
In this viewpoint,
two sides of Eqs.~(\ref{pi}) and (\ref{pi_berry}) are connected through an $n$-fold suspension isomorphism.
However, in our opinion, the lattice construction becomes vague and complicated after $n$ times of suspension.
By comparison,
our construction explained below Eq.~(\ref{pik}) is more explicit and isotropic in the sense that we treat each spatial direction on equal footing as shown in FIG.~\ref{sphere}.
This advantage will be useful in designing a multi-variable modulating texture in a lattice Hamiltonian in Inv$_d^G$.

The non-contractibility of a mapping in $[S^n,\mathcal{P}(\text{Inv}_{m}^G)]_0$ implies a nontrivial gappability index $\mathcal{I}_G\leq n+1$ possessed by at least one of the non-invertible Hamiltonians, e.g., {those} in gapless phases~\cite{Yao:2022vh}, which are {in the region enclosed by} the image of the non-contractible mapping.
However,
to pinpoint these representatives will not be a simple task in practice,
although we expect that they must be surrounded by the non-contractible $S^n$ in our construction.
The numerical investigation of them is left for future work.

Finally,
our construction of crystalline topological phases is expected to be generalizable towards nontrivial family of Hamiltonians,
and its physical consequences {will also be of} future interest.

\section{Acknowledgments}
We thank  Takamasa Ando, Masaki Oshikawa, and Ken Shiozaki for useful discussions.
Y.~Y. thanks the sponsorship from Yangyang Development Fund and Xiaomi Young Scholars Program.
This work was supported in part by JSPS KAKENHI (Grant No.\ JP19K03680) and JST CREST (Grant No.\ JPMJCR19T2).

\appendix*
\section{$Q_8$-symmetric path between distinct $\mathbb{Z}_2\times\mathbb{Z}_2$ SPT phases}\label{appendix}
The following two spin-1/2 chains
\begin{eqnarray}
H_1&=&\sum_n\frac{1}{2}[1+(-1)^n]\vec{S}_n\cdot\vec{S}_{n+1},\\
H_2&=&\sum_n\frac{1}{2}[1-(-1)^n]\vec{S}_n\cdot\vec{S}_{n+1},
\end{eqnarray}
belong to the two distinct $\mathbb{Z}_2\times\mathbb{Z}_2$ SPT phases,
where the onsite symmetry $\mathbb{Z}_2\times\mathbb{Z}_2$ is generated by $\pi$-rotation symmetries:
\begin{eqnarray}
R_z\equiv \prod_ni\sigma^z_n;\,\,R_x\equiv\prod_ni\sigma^x_n,
\end{eqnarray}
where $\vec{\sigma}_n$ represent the Pauli matrices acting on the $n$-th spin-1/2.
$\mathbb{Z}_2\times\mathbb{Z}_2$ is a subgroup of the conventional SO(3) spin-rotation symmetry,
so it also inherits the $\mathbb{Z}_2$-gauge structure when SO(3) is treated as SU(2) symmetry.
On each site,
it becomes the quaternion group $Q_8$:
\begin{eqnarray}
Q_8=\left\{1,-1,i\sigma^z,i\sigma^x,i\sigma^y,-i\sigma^z,-i\sigma^x,-i\sigma^y\right\}\subset \text{SU}(2),\nonumber
\end{eqnarray}
and $\mathbb{Z}_2\times\mathbb{Z}_2=Q_8/\mathbb{Z}_2$ with $\mathbb{Z}_2=\{1,-1\}$ paralleling $\text{SO}(3)=\text{SU}(2)/\mathbb{Z}_2$.
We associate these group elements with their physical operators denoted as
\begin{eqnarray}
i\sigma^z\leftrightarrow r^z;i\sigma^x\leftrightarrow r^x.
\end{eqnarray}

Our goal in this Appendix is to give an atomic $Q_8$-symmetric Hamiltonian $E$ which realizes a $Q_8$-symmetric adiabatic path
\begin{eqnarray}\label{stable}
H_1\h E\Leftrightarrow H_2\h E,
\end{eqnarray}
while there is no $\mathbb{Z}_2\times\mathbb{Z}_2$-symmetric $E$ to do this job.

To construct $E$,
we first build its local Hilbert space generated by two-flavor boson $b$ and $d$:
\begin{eqnarray}
\{|0\rangle,b^\dagger|0\rangle,d^\dagger|0\rangle,(b^\dagger )^2|0\rangle,b^\dagger d^\dagger|0\rangle, (d^\dagger)^2|0\rangle,\cdots\}.
\end{eqnarray}
We define an onsite $Q_8$ symmetry on it by
\begin{eqnarray}
&&r^x|0\rangle=|0\rangle, \qquad r^z|0\rangle=|0\rangle;\nonumber\\
&&\left(\begin{array}{c}r^zb^\dagger (r^z)^{-1}\\r^zd^\dagger (r^z)^{-1}\end{array}\right)=i\sigma_z\!\left(\begin{array}{c}b^\dagger \\d^\dagger\end{array}\right)=\left(\begin{array}{c}ib^\dagger \\-id^\dagger\end{array}\right),\\
&&\left(\begin{array}{c}r^xb^\dagger (r^x)^{-1}\\r^xd^\dagger (r^x)^{-1}\end{array}\right)=i\sigma_x\!\left(\begin{array}{c}b^\dagger \\d^\dagger\end{array}\right)=\left(\begin{array}{c}id^\dagger \\ib^\dagger\end{array}\right),\nonumber
\end{eqnarray}
i.e., the vacuum state $|0\rangle$ forms a singlet, and the one-particle states form a fundamental representation.
It should be noted that $\mathbb{Z}_2\in Q_8$, which is generated by $(r^z)^2$, can no longer be regarded as a gauge structure, since, e.g.,
\begin{eqnarray}
(r^z)^2[|0\rangle+b^\dagger|0\rangle]=|0\rangle-b^\dagger|0\rangle\neq e^{i\phi}[|0\rangle+b^\dagger|0\rangle]
\end{eqnarray}
for any phase factor $\phi$.

Let us consider a four-site $b$-$d$ bosonic system; $\{b_n,d_n\}_{n=1,2,3,4}$ and we still use $|0\rangle$ to label the empty state for the entire system.
We have the following two states:
\begin{eqnarray}
|2020\rangle&\equiv&\left(b^\dagger_1d^\dagger_1\right)\left(b^\dagger_3d^\dagger_3\right)|0\rangle;\nonumber\\
|(11)(11)\rangle&\equiv&\left(b^\dagger_1d^\dagger_2-d^\dagger_1b^\dagger_2\right)\left(b^\dagger_3d^\dagger_4-d^\dagger_3b^\dagger_4\right)|0\rangle,\nonumber
\end{eqnarray}
where we have used the particle number to label the states.
The following two Hamiltonians have $|2020\rangle$ and $|(11)(11)\rangle$ as their unique ground states, respectively:
\begin{eqnarray}
h_{2020}&=&-|2020\rangle\langle2020|;\\
h_{(11)(11)}&=&-|(11)(11)\rangle\langle(11)(11)|.
\end{eqnarray}

It is also clear that these two states have the same eigenvalue $+1$ for both $Q_8$ generators $r^x$ and $r^z$,
so there is an adiabatic path connecting $h_{2020}$ and $h_{(11)(11)}$, e.g.,
\begin{eqnarray}
h(s)&=&-\left[\cos\!\left(\frac{s\pi}{2}\right)|2020\rangle+\sin\!\left(\frac{s\pi}{2}\right)|(11)(11)\rangle\right] \nonumber\\
     &&  \times \left[\cos\!\left(\frac{s\pi}{2}\right)\langle 2020|+\sin\!\left(\frac{s\pi}{2}\right)\langle(11)(11)|\right]
     \nonumber\\&&
\end{eqnarray}
with $s\in[0,1]$.

Furthermore,
the following Hamiltonian has exactly the same unique ground state as $h_{(11)(11)}$:
\begin{eqnarray}
h'_{(11)(11)}&=&\vec{s}_1\cdot\vec{s}_2+\vec{s}_3\cdot\vec{s}_4 \nonumber\\
&&+10000\sum_{n=1}^4\left(b_n^\dagger b_n+d_n^\dagger d_n-1\right)^2,
\end{eqnarray}
with
\begin{equation}
\vec{s}_n\equiv
\begin{pmatrix}b_n^\dagger & d_n^\dagger\end{pmatrix}
\frac{\vec{\sigma}}{2}\begin{pmatrix} b_n\\ d_n\end{pmatrix}.
\end{equation}
There is an adiabatic path from $h_{(11)(11)}$ to $h'_{(11)(11)}$ by simply adjusting the energies of the ground state and the excited states without closing the gap between the ground state and the lowest excited state;
see Eq.~(\ref{moving}).

Similarly,
the $Q_8$-symmetric Hamiltonian
\begin{eqnarray}
h'_{2020}\equiv 10000 \sum_{n=1}^4\left\{[b_{n}^\dagger b_{n}+(-1)^n]^2+[d_{n}^\dagger d_{n}+(-1)^n]^2\right\}\nonumber
\end{eqnarray}
also has the same unique gapped ground state as $h_{2020}$,
so there is an adiabatic path from $h'_{2020}$ to $h_{2020}$.

Therefore,
together with $h(s)$,
we have obtained an adiabatic path from $h'_{2020}$ to $h'_{(11)(11)}$.

Repeating the same argument,
we also have an adiabatic path connecting $h'_{(11)(11)}$ and $h'_{0202}$ defined by
\begin{eqnarray}
h'_{0202}\equiv 10000 \sum_{n=1}^4\left\{[b_{n}^\dagger b_{n}-(-1)^n]^2+[d_{n}^\dagger d_{n}-(-1)^n]^2\right\},\nonumber
\end{eqnarray}
which is equivalent to relabelling the sites backward in the earlier path from $h'_{2020}$ to $h'_{(11)(11)}$.

Then we generalize them to a chain of $b$-$d$ bosons:
\begin{eqnarray}
E\equiv 10000 \sum_{n\in\mathbb{Z}}\left\{[b_{n}^\dagger b_{n}+(-1)^n]^2+[d_{n}^\dagger d_{n}+(-1)^n]^2\right\},\nonumber
\end{eqnarray}
which is to be shown as the atomic Hamiltonian to realize Eq.~(\ref{stable}) as follows.

We also define
\begin{eqnarray}
E'&\equiv&\sum_n\frac{1}{2}[1+(-1)^n]\vec{s}_n\cdot\vec{s}_{n+1} \nonumber\\
&&+10000\sum_n\left(b_n^\dagger b_n+d_n^\dagger d_n-1\right)^2,\\
E''&\equiv&\sum_n\frac{1}{2}[1-(-1)^n]\vec{s}_n\cdot\vec{s}_{n+1} \nonumber\\
&&+10000\sum_n\left(b_n^\dagger b_n+d_n^\dagger d_n-1\right)^2.
\end{eqnarray}
By the above four-site paths,
we obtain a path from $E$ to $E'$ and a path from $E$ to $E''$,
thereby trivially
\begin{eqnarray}
H_1\h E\Leftrightarrow H_1\h E';\\
H_2\h E\Leftrightarrow H_2\h E''.
\end{eqnarray}
Now $H_1\h E'$ can be seen as two decoupled dimerized spin-1/2 chains at the energy scale below $1000$,
so there is an SU(2)-symmetric adiabatic path from it to a ladder Hamiltonian $H_\text{ladder}$ through an intermediate plaquette Hamiltonian $H_\text{plaq}$,
and $H_2\h E''$ is also adiabatically connected to the same $H_\text{ladder}$ through the other plaquette Hamiltonian $H'_\text{plaq}$ (which differs from the earlier $H_\text{plaq}$ by a lattice translation).
These are adiabatic paths along which the many-body gap does not close~\cite{Marshall:1955tw,Lieb:1962wk,Lieb:1989wr,Chitov:2008aa}.
Consequently,
we have proven Eq.~(\ref{stable}) by explicitly constructing the $Q_8$-symmetric path,
\begin{widetext}
\begin{eqnarray}
H_1\h E\Leftrightarrow H_1\h E'\Leftrightarrow H_\text{plaq}\Leftrightarrow H_\text{ladder}
\Leftrightarrow H'_\text{plaq}\Leftrightarrow H_2\h E''\Leftrightarrow H_2\h E.
\end{eqnarray}
\end{widetext}


\begin{thebibliography}{76}%
\makeatletter
\providecommand \@ifxundefined [1]{%
 \@ifx{#1\undefined}
}%
\providecommand \@ifnum [1]{%
 \ifnum #1\expandafter \@firstoftwo
 \else \expandafter \@secondoftwo
 \fi
}%
\providecommand \@ifx [1]{%
 \ifx #1\expandafter \@firstoftwo
 \else \expandafter \@secondoftwo
 \fi
}%
\providecommand \natexlab [1]{#1}%
\providecommand \enquote  [1]{``#1''}%
\providecommand \bibnamefont  [1]{#1}%
\providecommand \bibfnamefont [1]{#1}%
\providecommand \citenamefont [1]{#1}%
\providecommand \href@noop [0]{\@secondoftwo}%
\providecommand \href [0]{\begingroup \@sanitize@url \@href}%
\providecommand \@href[1]{\@@startlink{#1}\@@href}%
\providecommand \@@href[1]{\endgroup#1\@@endlink}%
\providecommand \@sanitize@url [0]{\catcode `\\12\catcode `\$12\catcode
  `\&12\catcode `\#12\catcode `\^12\catcode `\_12\catcode `\%12\relax}%
\providecommand \@@startlink[1]{}%
\providecommand \@@endlink[0]{}%
\providecommand \url  [0]{\begingroup\@sanitize@url \@url }%
\providecommand \@url [1]{\endgroup\@href {#1}{\urlprefix }}%
\providecommand \urlprefix  [0]{URL }%
\providecommand \Eprint [0]{\href }%
\providecommand \doibase [0]{http://dx.doi.org/}%
\providecommand \selectlanguage [0]{\@gobble}%
\providecommand \bibinfo  [0]{\@secondoftwo}%
\providecommand \bibfield  [0]{\@secondoftwo}%
\providecommand \translation [1]{[#1]}%
\providecommand \BibitemOpen [0]{}%
\providecommand \bibitemStop [0]{}%
\providecommand \bibitemNoStop [0]{.\EOS\space}%
\providecommand \EOS [0]{\spacefactor3000\relax}%
\providecommand \BibitemShut  [1]{\csname bibitem#1\endcsname}%
\let\auto@bib@innerbib\@empty
\bibitem [{\citenamefont {Gu}\ and\ \citenamefont {Wen}(2009)}]{Gu:2009aa}%
  \BibitemOpen
  \bibfield  {author} {\bibinfo {author} {\bibfnamefont {Zheng-Cheng}\
  \bibnamefont {Gu}}\ and\ \bibinfo {author} {\bibfnamefont {Xiao-Gang}\
  \bibnamefont {Wen}},\ }\bibfield  {title} {\enquote {\bibinfo {title}
  {Tensor-entanglement-filtering renormalization approach and
  symmetry-protected topological order},}\ }\href
  {https://link.aps.org/doi/10.1103/PhysRevB.80.155131} {\bibfield  {journal}
  {\bibinfo  {journal} {Phys. Rev. B}\ }\textbf {\bibinfo {volume} {80}},\
  \bibinfo {pages} {155131--} (\bibinfo {year} {2009})}\BibitemShut {NoStop}%
\bibitem [{\citenamefont {Kitaev}()}]{Kitaev:2011aa}%
  \BibitemOpen
  \bibfield  {author} {\bibinfo {author} {\bibfnamefont {A}~\bibnamefont
  {Kitaev}},\ }\bibfield  {title} {\enquote {\bibinfo {title} {Toward a
  topological classification of manybody quantum states with short-range
  entanglement in: Topological quantum computing workshop},}\ }\href
  {http://scgp.stonybrook.edu/video_portal/video.php?id=1835} {\bibinfo
  {journal} {\textit{Toward a topological classification of manybody quantum
  states with short-range entanglement}, Topological Quantum Computing
  Workshop, Simons Center for Geometry and Physics (Stony Brook University,
  Stony Brook, New York, 2011)}\ }\BibitemShut {NoStop}%
\bibitem [{\citenamefont {Kitaev}(2013)}]{Kitaev:2013aa}%
  \BibitemOpen
\bibfield  {journal} {  }\bibfield  {author} {\bibinfo {author} {\bibfnamefont
  {Alexei}\ \bibnamefont {Kitaev}},\ }\bibfield  {title} {\enquote {\bibinfo
  {title} {On the classification of short-range entangled states},}\ }\href
  {https://scgp.stonybrook.edu/archives/7874} {\bibfield  {journal} {\bibinfo
  {journal} {\textit{On the classification of short-range entangled states},
  Simons Center for Geometry and Physics, Stony Brook University Stony Brook,
  NY}\ } (\bibinfo {year} {2013})}\BibitemShut {NoStop}%
\bibitem [{\citenamefont {Chen}\ \emph {et~al.}(2010)\citenamefont {Chen},
  \citenamefont {Gu},\ and\ \citenamefont {Wen}}]{Chen:2010aa}%
  \BibitemOpen
  \bibfield  {author} {\bibinfo {author} {\bibfnamefont {Xie}\ \bibnamefont
  {Chen}}, \bibinfo {author} {\bibfnamefont {Zheng-Cheng}\ \bibnamefont {Gu}},
  \ and\ \bibinfo {author} {\bibfnamefont {Xiao-Gang}\ \bibnamefont {Wen}},\
  }\bibfield  {title} {\enquote {\bibinfo {title} {Local unitary
  transformation, long-range quantum entanglement, wave function
  renormalization, and topological order},}\ }\href
  {https://doi.org/10.1103/PhysRevB.82.155138} {\bibfield  {journal} {\bibinfo
  {journal} {Phys. Rev. B}\ }\textbf {\bibinfo {volume} {82}},\ \bibinfo
  {pages} {155138} (\bibinfo {year} {2010})}\BibitemShut {NoStop}%
\bibitem [{\citenamefont {Pollmann}\ \emph {et~al.}(2012)\citenamefont
  {Pollmann}, \citenamefont {Berg}, \citenamefont {Turner},\ and\ \citenamefont
  {Oshikawa}}]{Pollmann:2012aa}%
  \BibitemOpen
  \bibfield  {author} {\bibinfo {author} {\bibfnamefont {Frank}\ \bibnamefont
  {Pollmann}}, \bibinfo {author} {\bibfnamefont {Erez}\ \bibnamefont {Berg}},
  \bibinfo {author} {\bibfnamefont {Ari~M.}\ \bibnamefont {Turner}}, \ and\
  \bibinfo {author} {\bibfnamefont {Masaki}\ \bibnamefont {Oshikawa}},\
  }\bibfield  {title} {\enquote {\bibinfo {title} {Symmetry protection of
  topological phases in one-dimensional quantum spin systems},}\ }\href
  {https://link.aps.org/doi/10.1103/PhysRevB.85.075125} {\bibfield  {journal}
  {\bibinfo  {journal} {Phys. Rev. B}\ }\textbf {\bibinfo {volume} {85}},\
  \bibinfo {pages} {075125--} (\bibinfo {year} {2012})}\BibitemShut {NoStop}%
\bibitem [{\citenamefont {Wen}(2013)}]{Wen:TOreview2013}%
  \BibitemOpen
  \bibfield  {author} {\bibinfo {author} {\bibfnamefont {Xiao-Gang}\
  \bibnamefont {Wen}},\ }\bibfield  {title} {\enquote {\bibinfo {title}
  {Topological order: From long-range entangled quantum matter to a unified
  origin of light and electrons},}\ }\href {\doibase
  http://dx.doi.org/10.1155/2013/198710} {\bibfield  {journal} {\bibinfo
  {journal} {ISRN Condensed Matter Physics}\ }\textbf {\bibinfo {volume}
  {2013}},\ \bibinfo {pages} {198710} (\bibinfo {year} {2013})}\BibitemShut
  {NoStop}%
\bibitem [{\citenamefont {Kitaev}(2001)}]{Kitaev:2001aa}%
  \BibitemOpen
  \bibfield  {author} {\bibinfo {author} {\bibfnamefont {A~Yu}\ \bibnamefont
  {Kitaev}},\ }\bibfield  {title} {\enquote {\bibinfo {title} {Unpaired
  majorana fermions in quantum wires},}\ }\href
  {https://doi.org/10.1070/1063-7869/44/10S/S29} {\bibfield  {journal}
  {\bibinfo  {journal} {Phys.-Usp.}\ }\textbf {\bibinfo {volume} {44}},\
  \bibinfo {pages} {131} (\bibinfo {year} {2001})}\BibitemShut {NoStop}%
\bibitem [{\citenamefont {Schnyder}\ \emph {et~al.}(2008)\citenamefont
  {Schnyder}, \citenamefont {Ryu}, \citenamefont {Furusaki},\ and\
  \citenamefont {Ludwig}}]{Schnyder:2008aa}%
  \BibitemOpen
  \bibfield  {author} {\bibinfo {author} {\bibfnamefont {Andreas~P.}\
  \bibnamefont {Schnyder}}, \bibinfo {author} {\bibfnamefont {Shinsei}\
  \bibnamefont {Ryu}}, \bibinfo {author} {\bibfnamefont {Akira}\ \bibnamefont
  {Furusaki}}, \ and\ \bibinfo {author} {\bibfnamefont {Andreas W.~W.}\
  \bibnamefont {Ludwig}},\ }\bibfield  {title} {\enquote {\bibinfo {title}
  {Classification of topological insulators and superconductors in three
  spatial dimensions},}\ }\href
  {https://link.aps.org/doi/10.1103/PhysRevB.78.195125} {\bibfield  {journal}
  {\bibinfo  {journal} {Phys. Rev. B}\ }\textbf {\bibinfo {volume} {78}},\
  \bibinfo {pages} {195125--} (\bibinfo {year} {2008})}\BibitemShut {NoStop}%
\bibitem [{\citenamefont {Ryu}\ \emph {et~al.}(2010)\citenamefont {Ryu},
  \citenamefont {Schnyder}, \citenamefont {Furusaki},\ and\ \citenamefont
  {Ludwig}}]{Ryu2010}%
  \BibitemOpen
  \bibfield  {author} {\bibinfo {author} {\bibfnamefont {Shinsei}\ \bibnamefont
  {Ryu}}, \bibinfo {author} {\bibfnamefont {Andreas~P}\ \bibnamefont
  {Schnyder}}, \bibinfo {author} {\bibfnamefont {Akira}\ \bibnamefont
  {Furusaki}}, \ and\ \bibinfo {author} {\bibfnamefont {Andreas W~W}\
  \bibnamefont {Ludwig}},\ }\bibfield  {title} {\enquote {\bibinfo {title}
  {Topological insulators and superconductors: tenfold way and dimensional
  hierarchy},}\ }\href {\doibase 10.1088/1367-2630/12/6/065010} {\bibfield
  {journal} {\bibinfo  {journal} {New Journal of Physics}\ }\textbf {\bibinfo
  {volume} {12}},\ \bibinfo {pages} {065010} (\bibinfo {year}
  {2010})}\BibitemShut {NoStop}%
\bibitem [{\citenamefont {Kitaev}(2009)}]{Kitaev2009}%
  \BibitemOpen
  \bibfield  {author} {\bibinfo {author} {\bibfnamefont {Alexei}\ \bibnamefont
  {Kitaev}},\ }\bibfield  {title} {\enquote {\bibinfo {title} {Periodic table
  for topological insulators and superconductors},}\ }\href {\doibase
  10.1063/1.3149495} {\bibfield  {journal} {\bibinfo  {journal} {AIP Conference
  Proceedings}\ }\textbf {\bibinfo {volume} {1134}},\ \bibinfo {pages} {22--30}
  (\bibinfo {year} {2009})},\ \Eprint
  {http://arxiv.org/abs/https://aip.scitation.org/doi/pdf/10.1063/1.3149495}
  {https://aip.scitation.org/doi/pdf/10.1063/1.3149495} \BibitemShut {NoStop}%
\bibitem [{\citenamefont {Kapustin}(2014)}]{Kapustin:2014ac}%
  \BibitemOpen
  \bibfield  {author} {\bibinfo {author} {\bibfnamefont {Anton}\ \bibnamefont
  {Kapustin}},\ }\bibfield  {title} {\enquote {\bibinfo {title} {Symmetry
  protected topological phases, anomalies, and cobordisms: beyond group
  cohomology},}\ }\href {https://arxiv.org/abs/1403.1467} {\bibfield  {journal}
  {\bibinfo  {journal} {arXiv preprint arXiv:1403.1467}\ } (\bibinfo {year}
  {2014})}\BibitemShut {NoStop}%
\bibitem [{\citenamefont {Kapustin}\ \emph {et~al.}(2015)\citenamefont
  {Kapustin}, \citenamefont {Thorngren}, \citenamefont {Turzillo},\ and\
  \citenamefont {Wang}}]{Kapustin:2015aa}%
  \BibitemOpen
  \bibfield  {author} {\bibinfo {author} {\bibfnamefont {Anton}\ \bibnamefont
  {Kapustin}}, \bibinfo {author} {\bibfnamefont {Ryan}\ \bibnamefont
  {Thorngren}}, \bibinfo {author} {\bibfnamefont {Alex}\ \bibnamefont
  {Turzillo}}, \ and\ \bibinfo {author} {\bibfnamefont {Zitao}\ \bibnamefont
  {Wang}},\ }\bibfield  {title} {\enquote {\bibinfo {title} {Fermionic symmetry
  protected topological phases and cobordisms},}\ }\href
  {https://doi.org/10.1007/JHEP12%282015%29052} {\bibfield  {journal} {\bibinfo
   {journal} {JHEP}\ }\textbf {\bibinfo {volume} {2015}},\ \bibinfo {pages}
  {1--21} (\bibinfo {year} {2015})}\BibitemShut {NoStop}%
\bibitem [{\citenamefont {Gaiotto}\ and\ \citenamefont
  {Johnson-Freyd}(2019)}]{Gaiotto:2019aa}%
  \BibitemOpen
  \bibfield  {author} {\bibinfo {author} {\bibfnamefont {Davide}\ \bibnamefont
  {Gaiotto}}\ and\ \bibinfo {author} {\bibfnamefont {Theo}\ \bibnamefont
  {Johnson-Freyd}},\ }\bibfield  {title} {\enquote {\bibinfo {title} {Symmetry
  protected topological phases and generalized cohomology},}\ }\href
  {https://doi.org/10.1007/JHEP05%282019%29007} {\bibfield  {journal} {\bibinfo
   {journal} {JHEP}\ }\textbf {\bibinfo {volume} {2019}},\ \bibinfo {pages}
  {1--36} (\bibinfo {year} {2019})}\BibitemShut {NoStop}%
\bibitem [{\citenamefont {'t~Hooft}\ \emph {et~al.}(1980)\citenamefont
  {'t~Hooft}, \citenamefont {Itzykson}, \citenamefont {Jaffe}, \citenamefont
  {Lehmann}, \citenamefont {Mitter}, \citenamefont {Singer},\ and\
  \citenamefont {Stora}}]{tHooft:1980xss}%
  \BibitemOpen
  \bibfield  {author} {\bibinfo {author} {\bibfnamefont {Gerard}\ \bibnamefont
  {'t~Hooft}}, \bibinfo {author} {\bibfnamefont {C.}~\bibnamefont {Itzykson}},
  \bibinfo {author} {\bibfnamefont {A.}~\bibnamefont {Jaffe}}, \bibinfo
  {author} {\bibfnamefont {H.}~\bibnamefont {Lehmann}}, \bibinfo {author}
  {\bibfnamefont {P.~K.}\ \bibnamefont {Mitter}}, \bibinfo {author}
  {\bibfnamefont {I.~M.}\ \bibnamefont {Singer}}, \ and\ \bibinfo {author}
  {\bibfnamefont {R.}~\bibnamefont {Stora}},\ }\bibfield  {title} {\enquote
  {\bibinfo {title} {{Recent Developments in Gauge Theories. Proceedings, Nato
  Advanced Study Institute, Cargese, France, August 26 - September 8, 1979}},}\
  }\href {\doibase 10.1007/978-1-4684-7571-5} {\bibfield  {journal} {\bibinfo
  {journal} {NATO Sci. Ser. B}\ }\textbf {\bibinfo {volume} {59}},\ \bibinfo
  {pages} {pp.1--438} (\bibinfo {year} {1980})}\BibitemShut {NoStop}%
\bibitem [{\citenamefont {Seiberg}\ and\ \citenamefont
  {Witten}(2016)}]{Seiberg:2016aa}%
  \BibitemOpen
  \bibfield  {author} {\bibinfo {author} {\bibfnamefont {Nathan}\ \bibnamefont
  {Seiberg}}\ and\ \bibinfo {author} {\bibfnamefont {Edward}\ \bibnamefont
  {Witten}},\ }\bibfield  {title} {\enquote {\bibinfo {title} {Gapped boundary
  phases of topological insulators via weak coupling},}\ }\href
  {https://academic.oup.com/ptep/article/2016/12/12C101/2624087} {\bibfield
  {journal} {\bibinfo  {journal} {Progr. Theor. Phys.}\ }\textbf {\bibinfo
  {volume} {2016}} (\bibinfo {year} {2016})}\BibitemShut {NoStop}%
\bibitem [{\citenamefont {Witten}(2016{\natexlab{a}})}]{Witten:2016aa}%
  \BibitemOpen
  \bibfield  {author} {\bibinfo {author} {\bibfnamefont {Edward}\ \bibnamefont
  {Witten}},\ }\bibfield  {title} {\enquote {\bibinfo {title} {Fermion path
  integrals and topological phases},}\ }\href
  {https://link.aps.org/doi/10.1103/RevModPhys.88.035001} {\bibfield  {journal}
  {\bibinfo  {journal} {Rev. Mod. Phys.}\ }\textbf {\bibinfo {volume} {88}},\
  \bibinfo {pages} {035001--} (\bibinfo {year}
  {2016}{\natexlab{a}})}\BibitemShut {NoStop}%
\bibitem [{\citenamefont {Witten}(1985)}]{Witten:1985aa}%
  \BibitemOpen
  \bibfield  {author} {\bibinfo {author} {\bibfnamefont {Edward}\ \bibnamefont
  {Witten}},\ }\bibfield  {title} {\enquote {\bibinfo {title} {Global
  gravitational anomalies},}\ }\href
  {https://projecteuclid.org/journals/communications-in-mathematical-physics/volume-100/issue-2/Global-gravitational-anomalies/cmp/1103943444.full}
  {\bibfield  {journal} {\bibinfo  {journal} {Comm. Math. Phys.}\ }\textbf
  {\bibinfo {volume} {100}},\ \bibinfo {pages} {197--229} (\bibinfo {year}
  {1985})}\BibitemShut {NoStop}%
\bibitem [{\citenamefont {Teo}\ and\ \citenamefont {Kane}(2010)}]{Teo:2010aa}%
  \BibitemOpen
  \bibfield  {author} {\bibinfo {author} {\bibfnamefont {Jeffrey~CY}\
  \bibnamefont {Teo}}\ and\ \bibinfo {author} {\bibfnamefont {Charles~L}\
  \bibnamefont {Kane}},\ }\bibfield  {title} {\enquote {\bibinfo {title}
  {Topological defects and gapless modes in insulators and superconductors},}\
  }\href {https://doi.org/10.1103/PhysRevB.82.115120} {\bibfield  {journal}
  {\bibinfo  {journal} {Phys. Rev. B}\ }\textbf {\bibinfo {volume} {82}},\
  \bibinfo {pages} {115120} (\bibinfo {year} {2010})}\BibitemShut {NoStop}%
\bibitem [{\citenamefont {Hsin}\ \emph {et~al.}(2020)\citenamefont {Hsin},
  \citenamefont {Kapustin},\ and\ \citenamefont {Thorngren}}]{Hsin:2020aa}%
  \BibitemOpen
  \bibfield  {author} {\bibinfo {author} {\bibfnamefont {Po-Shen}\ \bibnamefont
  {Hsin}}, \bibinfo {author} {\bibfnamefont {Anton}\ \bibnamefont {Kapustin}},
  \ and\ \bibinfo {author} {\bibfnamefont {Ryan}\ \bibnamefont {Thorngren}},\
  }\bibfield  {title} {\enquote {\bibinfo {title} {Berry phase in quantum field
  theory: Diabolical points and boundary phenomena},}\ }\href
  {https://doi.org/10.1103/PhysRevB.102.245113} {\bibfield  {journal} {\bibinfo
   {journal} {Phys. Rev. B}\ }\textbf {\bibinfo {volume} {102}},\ \bibinfo
  {pages} {245113} (\bibinfo {year} {2020})}\BibitemShut {NoStop}%
\bibitem [{\citenamefont {Cordova}\ \emph
  {et~al.}(2020{\natexlab{a}})\citenamefont {Cordova}, \citenamefont {Freed},
  \citenamefont {Lam},\ and\ \citenamefont {Seiberg}}]{Cordova:2020aa}%
  \BibitemOpen
  \bibfield  {author} {\bibinfo {author} {\bibfnamefont {Clay}\ \bibnamefont
  {Cordova}}, \bibinfo {author} {\bibfnamefont {Daniel}\ \bibnamefont {Freed}},
  \bibinfo {author} {\bibfnamefont {Ho~Tat}\ \bibnamefont {Lam}}, \ and\
  \bibinfo {author} {\bibfnamefont {Nathan}\ \bibnamefont {Seiberg}},\
  }\bibfield  {title} {\enquote {\bibinfo {title} {Anomalies in the space of
  coupling constants and their dynamical applications i},}\ }\href
  {https://doi.org/10.21468/SciPostPhys.8.1.001} {\bibfield  {journal}
  {\bibinfo  {journal} {SciPost Physics}\ }\textbf {\bibinfo {volume} {8}},\
  \bibinfo {pages} {1} (\bibinfo {year} {2020}{\natexlab{a}})}\BibitemShut
  {NoStop}%
\bibitem [{\citenamefont {Cordova}\ \emph
  {et~al.}(2020{\natexlab{b}})\citenamefont {Cordova}, \citenamefont {Freed},
  \citenamefont {Lam},\ and\ \citenamefont {Seiberg}}]{Cordova:2020ab}%
  \BibitemOpen
  \bibfield  {author} {\bibinfo {author} {\bibfnamefont {Clay}\ \bibnamefont
  {Cordova}}, \bibinfo {author} {\bibfnamefont {Daniel~S}\ \bibnamefont
  {Freed}}, \bibinfo {author} {\bibfnamefont {Ho~Tat}\ \bibnamefont {Lam}}, \
  and\ \bibinfo {author} {\bibfnamefont {Nathan}\ \bibnamefont {Seiberg}},\
  }\bibfield  {title} {\enquote {\bibinfo {title} {Anomalies in the space of
  coupling constants and their dynamical applications ii},}\ }\href
  {https://doi.org/10.21468/SciPostPhys.8.1.002} {\bibfield  {journal}
  {\bibinfo  {journal} {SciPost Physics}\ }\textbf {\bibinfo {volume} {8}},\
  \bibinfo {pages} {2} (\bibinfo {year} {2020}{\natexlab{b}})}\BibitemShut
  {NoStop}%
\bibitem [{\citenamefont {Kapustin}\ and\ \citenamefont
  {Spodyneiko}(2020{\natexlab{a}})}]{Kapustin:2020aa}%
  \BibitemOpen
  \bibfield  {author} {\bibinfo {author} {\bibfnamefont {Anton}\ \bibnamefont
  {Kapustin}}\ and\ \bibinfo {author} {\bibfnamefont {Lev}\ \bibnamefont
  {Spodyneiko}},\ }\bibfield  {title} {\enquote {\bibinfo {title}
  {Higher-dimensional generalizations of berry curvature},}\ }\href
  {https://doi.org/10.1103/PhysRevB.101.235130} {\bibfield  {journal} {\bibinfo
   {journal} {Phys. Rev. B}\ }\textbf {\bibinfo {volume} {101}},\ \bibinfo
  {pages} {235130} (\bibinfo {year} {2020}{\natexlab{a}})}\BibitemShut
  {NoStop}%
\bibitem [{\citenamefont {Wen}\ \emph {et~al.}(2023)\citenamefont {Wen},
  \citenamefont {Qi}, \citenamefont {Beaudry}, \citenamefont {Moreno},
  \citenamefont {Pflaum}, \citenamefont {Spiegel}, \citenamefont {Vishwanath},\
  and\ \citenamefont {Hermele}}]{Wen:2021aa}%
  \BibitemOpen
  \bibfield  {author} {\bibinfo {author} {\bibfnamefont {Xueda}\ \bibnamefont
  {Wen}}, \bibinfo {author} {\bibfnamefont {Marvin}\ \bibnamefont {Qi}},
  \bibinfo {author} {\bibfnamefont {Agn\`es}\ \bibnamefont {Beaudry}}, \bibinfo
  {author} {\bibfnamefont {Juan}\ \bibnamefont {Moreno}}, \bibinfo {author}
  {\bibfnamefont {Markus~J.}\ \bibnamefont {Pflaum}}, \bibinfo {author}
  {\bibfnamefont {Daniel}\ \bibnamefont {Spiegel}}, \bibinfo {author}
  {\bibfnamefont {Ashvin}\ \bibnamefont {Vishwanath}}, \ and\ \bibinfo {author}
  {\bibfnamefont {Michael}\ \bibnamefont {Hermele}},\ }\bibfield  {title}
  {\enquote {\bibinfo {title} {Flow of higher berry curvature and bulk-boundary
  correspondence in parametrized quantum systems},}\ }\href {\doibase
  10.1103/PhysRevB.108.125147} {\bibfield  {journal} {\bibinfo  {journal}
  {Phys. Rev. B}\ }\textbf {\bibinfo {volume} {108}},\ \bibinfo {pages}
  {125147} (\bibinfo {year} {2023})}\BibitemShut {NoStop}%
\bibitem [{\citenamefont {Kapustin}\ and\ \citenamefont
  {Spodyneiko}(2020{\natexlab{b}})}]{Kapustin:2020ab}%
  \BibitemOpen
  \bibfield  {author} {\bibinfo {author} {\bibfnamefont {Anton}\ \bibnamefont
  {Kapustin}}\ and\ \bibinfo {author} {\bibfnamefont {Lev}\ \bibnamefont
  {Spodyneiko}},\ }\bibfield  {title} {\enquote {\bibinfo {title}
  {Higher-dimensional generalizations of the thouless charge pump},}\ }\href
  {https://arxiv.org/abs/2003.09519} {\bibfield  {journal} {\bibinfo  {journal}
  {arXiv preprint arXiv:2003.09519}\ } (\bibinfo {year}
  {2020}{\natexlab{b}})}\BibitemShut {NoStop}%
\bibitem [{\citenamefont {Shiozaki}(2022)}]{Shiozaki:2021aa}%
  \BibitemOpen
  \bibfield  {author} {\bibinfo {author} {\bibfnamefont {Ken}\ \bibnamefont
  {Shiozaki}},\ }\bibfield  {title} {\enquote {\bibinfo {title} {Adiabatic
  cycles of quantum spin systems},}\ }\href {\doibase
  10.1103/PhysRevB.106.125108} {\bibfield  {journal} {\bibinfo  {journal}
  {Phys. Rev. B}\ }\textbf {\bibinfo {volume} {106}},\ \bibinfo {pages}
  {125108--} (\bibinfo {year} {2022})}\BibitemShut {NoStop}%
\bibitem [{\citenamefont {Ohyama}\ \emph {et~al.}(2023)\citenamefont {Ohyama},
  \citenamefont {Terashima},\ and\ \citenamefont {Shiozaki}}]{Ohyama:2023ty}%
  \BibitemOpen
  \bibfield  {author} {\bibinfo {author} {\bibfnamefont {Shuhei}\ \bibnamefont
  {Ohyama}}, \bibinfo {author} {\bibfnamefont {Yuji}\ \bibnamefont
  {Terashima}}, \ and\ \bibinfo {author} {\bibfnamefont {Ken}\ \bibnamefont
  {Shiozaki}},\ }\bibfield  {title} {\enquote {\bibinfo {title} {Discrete
  higher berry phases and matrix product states},}\ }\href
  {https://arxiv.org/abs/2303.04252} {\bibfield  {journal} {\bibinfo  {journal}
  {arXiv preprint arXiv:2303.04252}\ } (\bibinfo {year} {2023})}\BibitemShut
  {NoStop}%
\bibitem [{\citenamefont {Ohyama}\ \emph {et~al.}(2022)\citenamefont {Ohyama},
  \citenamefont {Shiozaki},\ and\ \citenamefont {Sato}}]{Ohyama:2022ts}%
  \BibitemOpen
  \bibfield  {author} {\bibinfo {author} {\bibfnamefont {Shuhei}\ \bibnamefont
  {Ohyama}}, \bibinfo {author} {\bibfnamefont {Ken}\ \bibnamefont {Shiozaki}},
  \ and\ \bibinfo {author} {\bibfnamefont {Masatoshi}\ \bibnamefont {Sato}},\
  }\bibfield  {title} {\enquote {\bibinfo {title} {Generalized thouless pumps
  in {\$}(1+1){\$}-dimensional interacting fermionic systems},}\ }\href
  {\doibase 10.1103/PhysRevB.106.165115} {\bibfield  {journal} {\bibinfo
  {journal} {Phys. Rev. B}\ }\textbf {\bibinfo {volume} {106}},\ \bibinfo
  {pages} {165115--} (\bibinfo {year} {2022})}\BibitemShut {NoStop}%
\bibitem [{\citenamefont {Yao}\ \emph {et~al.}(2022)\citenamefont {Yao},
  \citenamefont {Oshikawa},\ and\ \citenamefont {Furusaki}}]{Yao:2022vh}%
  \BibitemOpen
  \bibfield  {author} {\bibinfo {author} {\bibfnamefont {Yuan}\ \bibnamefont
  {Yao}}, \bibinfo {author} {\bibfnamefont {Masaki}\ \bibnamefont {Oshikawa}},
  \ and\ \bibinfo {author} {\bibfnamefont {Akira}\ \bibnamefont {Furusaki}},\
  }\bibfield  {title} {\enquote {\bibinfo {title} {Gappability index for
  quantum many-body systems},}\ }\href {\doibase
  10.1103/PhysRevLett.129.017204} {\bibfield  {journal} {\bibinfo  {journal}
  {Phys. Rev. Lett.}\ }\textbf {\bibinfo {volume} {129}},\ \bibinfo {pages}
  {017204--} (\bibinfo {year} {2022})}\BibitemShut {NoStop}%
\bibitem [{\citenamefont {Fu}(2011)}]{Fu:2011wr}%
  \BibitemOpen
  \bibfield  {author} {\bibinfo {author} {\bibfnamefont {Liang}\ \bibnamefont
  {Fu}},\ }\bibfield  {title} {\enquote {\bibinfo {title} {Topological
  crystalline insulators},}\ }\href {\doibase 10.1103/PhysRevLett.106.106802}
  {\bibfield  {journal} {\bibinfo  {journal} {Phys. Rev. Lett.}\ }\textbf
  {\bibinfo {volume} {106}},\ \bibinfo {pages} {106802--} (\bibinfo {year}
  {2011})}\BibitemShut {NoStop}%
\bibitem [{\citenamefont {Hsieh}\ \emph {et~al.}(2012)\citenamefont {Hsieh},
  \citenamefont {Lin}, \citenamefont {Liu}, \citenamefont {Duan}, \citenamefont
  {Bansil},\ and\ \citenamefont {Fu}}]{Hsieh:2012tq}%
  \BibitemOpen
  \bibfield  {author} {\bibinfo {author} {\bibfnamefont {Timothy~H.}\
  \bibnamefont {Hsieh}}, \bibinfo {author} {\bibfnamefont {Hsin}\ \bibnamefont
  {Lin}}, \bibinfo {author} {\bibfnamefont {Junwei}\ \bibnamefont {Liu}},
  \bibinfo {author} {\bibfnamefont {Wenhui}\ \bibnamefont {Duan}}, \bibinfo
  {author} {\bibfnamefont {Arun}\ \bibnamefont {Bansil}}, \ and\ \bibinfo
  {author} {\bibfnamefont {Liang}\ \bibnamefont {Fu}},\ }\bibfield  {title}
  {\enquote {\bibinfo {title} {Topological crystalline insulators in the snte
  material class},}\ }\href {\doibase 10.1038/ncomms1969} {\bibfield  {journal}
  {\bibinfo  {journal} {Nat. Commun}\ }\textbf {\bibinfo {volume} {3}},\
  \bibinfo {pages} {982} (\bibinfo {year} {2012})}\BibitemShut {NoStop}%
\bibitem [{\citenamefont {Fang}\ \emph {et~al.}(2012)\citenamefont {Fang},
  \citenamefont {Gilbert},\ and\ \citenamefont {Bernevig}}]{Fang:2012wp}%
  \BibitemOpen
  \bibfield  {author} {\bibinfo {author} {\bibfnamefont {Chen}\ \bibnamefont
  {Fang}}, \bibinfo {author} {\bibfnamefont {Matthew~J.}\ \bibnamefont
  {Gilbert}}, \ and\ \bibinfo {author} {\bibfnamefont {B.~Andrei}\ \bibnamefont
  {Bernevig}},\ }\bibfield  {title} {\enquote {\bibinfo {title} {Bulk
  topological invariants in noninteracting point group symmetric insulators},}\
  }\href {\doibase 10.1103/PhysRevB.86.115112} {\bibfield  {journal} {\bibinfo
  {journal} {Phys. Rev. B}\ }\textbf {\bibinfo {volume} {86}},\ \bibinfo
  {pages} {115112--} (\bibinfo {year} {2012})}\BibitemShut {NoStop}%
\bibitem [{\citenamefont {Chiu}\ \emph {et~al.}(2013)\citenamefont {Chiu},
  \citenamefont {Yao},\ and\ \citenamefont {Ryu}}]{Chiu:2013tk}%
  \BibitemOpen
  \bibfield  {author} {\bibinfo {author} {\bibfnamefont {Ching-Kai}\
  \bibnamefont {Chiu}}, \bibinfo {author} {\bibfnamefont {Hong}\ \bibnamefont
  {Yao}}, \ and\ \bibinfo {author} {\bibfnamefont {Shinsei}\ \bibnamefont
  {Ryu}},\ }\bibfield  {title} {\enquote {\bibinfo {title} {Classification of
  topological insulators and superconductors in the presence of reflection
  symmetry},}\ }\href {\doibase 10.1103/PhysRevB.88.075142} {\bibfield
  {journal} {\bibinfo  {journal} {Phys. Rev. B}\ }\textbf {\bibinfo {volume}
  {88}},\ \bibinfo {pages} {075142--} (\bibinfo {year} {2013})}\BibitemShut
  {NoStop}%
\bibitem [{\citenamefont {Morimoto}\ and\ \citenamefont
  {Furusaki}(2013)}]{Morimoto:2013wo}%
  \BibitemOpen
  \bibfield  {author} {\bibinfo {author} {\bibfnamefont {Takahiro}\
  \bibnamefont {Morimoto}}\ and\ \bibinfo {author} {\bibfnamefont {Akira}\
  \bibnamefont {Furusaki}},\ }\bibfield  {title} {\enquote {\bibinfo {title}
  {Topological classification with additional symmetries from clifford
  algebras},}\ }\href {\doibase 10.1103/PhysRevB.88.125129} {\bibfield
  {journal} {\bibinfo  {journal} {Phys. Rev. B}\ }\textbf {\bibinfo {volume}
  {88}},\ \bibinfo {pages} {125129--} (\bibinfo {year} {2013})}\BibitemShut
  {NoStop}%
\bibitem [{\citenamefont {Essin}\ and\ \citenamefont
  {Hermele}(2013)}]{Essin:2013wg}%
  \BibitemOpen
  \bibfield  {author} {\bibinfo {author} {\bibfnamefont {Andrew~M.}\
  \bibnamefont {Essin}}\ and\ \bibinfo {author} {\bibfnamefont {Michael}\
  \bibnamefont {Hermele}},\ }\bibfield  {title} {\enquote {\bibinfo {title}
  {Classifying fractionalization: Symmetry classification of gapped
  {\$}{\{}{$\backslash$}mathbb{\{}z{\}}{\}}{\_}{\{}2{\}}{\$} spin liquids in
  two dimensions},}\ }\href {\doibase 10.1103/PhysRevB.87.104406} {\bibfield
  {journal} {\bibinfo  {journal} {Phys. Rev. B}\ }\textbf {\bibinfo {volume}
  {87}},\ \bibinfo {pages} {104406--} (\bibinfo {year} {2013})}\BibitemShut
  {NoStop}%
\bibitem [{\citenamefont {Zhang}\ \emph {et~al.}(2013)\citenamefont {Zhang},
  \citenamefont {Kane},\ and\ \citenamefont {Mele}}]{Zhang:2013wh}%
  \BibitemOpen
  \bibfield  {author} {\bibinfo {author} {\bibfnamefont {Fan}\ \bibnamefont
  {Zhang}}, \bibinfo {author} {\bibfnamefont {C.~L.}\ \bibnamefont {Kane}}, \
  and\ \bibinfo {author} {\bibfnamefont {E.~J.}\ \bibnamefont {Mele}},\
  }\bibfield  {title} {\enquote {\bibinfo {title} {Topological mirror
  superconductivity},}\ }\href {\doibase 10.1103/PhysRevLett.111.056403}
  {\bibfield  {journal} {\bibinfo  {journal} {Phys. Rev. Lett.}\ }\textbf
  {\bibinfo {volume} {111}},\ \bibinfo {pages} {056403--} (\bibinfo {year}
  {2013})}\BibitemShut {NoStop}%
\bibitem [{\citenamefont {Shiozaki}\ and\ \citenamefont
  {Sato}(2014)}]{Shiozaki:2014wd}%
  \BibitemOpen
  \bibfield  {author} {\bibinfo {author} {\bibfnamefont {Ken}\ \bibnamefont
  {Shiozaki}}\ and\ \bibinfo {author} {\bibfnamefont {Masatoshi}\ \bibnamefont
  {Sato}},\ }\bibfield  {title} {\enquote {\bibinfo {title} {Topology of
  crystalline insulators and superconductors},}\ }\href {\doibase
  10.1103/PhysRevB.90.165114} {\bibfield  {journal} {\bibinfo  {journal} {Phys.
  Rev. B}\ }\textbf {\bibinfo {volume} {90}},\ \bibinfo {pages} {165114--}
  (\bibinfo {year} {2014})}\BibitemShut {NoStop}%
\bibitem [{\citenamefont {Hsieh}\ \emph {et~al.}(2014)\citenamefont {Hsieh},
  \citenamefont {Morimoto},\ and\ \citenamefont {Ryu}}]{Hsieh:2014us}%
  \BibitemOpen
  \bibfield  {author} {\bibinfo {author} {\bibfnamefont {Chang-Tse}\
  \bibnamefont {Hsieh}}, \bibinfo {author} {\bibfnamefont {Takahiro}\
  \bibnamefont {Morimoto}}, \ and\ \bibinfo {author} {\bibfnamefont {Shinsei}\
  \bibnamefont {Ryu}},\ }\bibfield  {title} {\enquote {\bibinfo {title} {Cpt
  theorem and classification of topological insulators and superconductors},}\
  }\href {\doibase 10.1103/PhysRevB.90.245111} {\bibfield  {journal} {\bibinfo
  {journal} {Phys. Rev. B}\ }\textbf {\bibinfo {volume} {90}},\ \bibinfo
  {pages} {245111--} (\bibinfo {year} {2014})}\BibitemShut {NoStop}%
\bibitem [{\citenamefont {Qi}\ and\ \citenamefont {Fu}(2015)}]{Qi:2015wj}%
  \BibitemOpen
  \bibfield  {author} {\bibinfo {author} {\bibfnamefont {Yang}\ \bibnamefont
  {Qi}}\ and\ \bibinfo {author} {\bibfnamefont {Liang}\ \bibnamefont {Fu}},\
  }\bibfield  {title} {\enquote {\bibinfo {title} {Anomalous crystal symmetry
  fractionalization on the surface of topological crystalline insulators},}\
  }\href {\doibase 10.1103/PhysRevLett.115.236801} {\bibfield  {journal}
  {\bibinfo  {journal} {Phys. Rev. B}\ }\textbf {\bibinfo {volume} {115}},\
  \bibinfo {pages} {236801--} (\bibinfo {year} {2015})}\BibitemShut {NoStop}%
\bibitem [{\citenamefont {Isobe}\ and\ \citenamefont
  {Fu}(2015)}]{Isobe:2015tf}%
  \BibitemOpen
  \bibfield  {author} {\bibinfo {author} {\bibfnamefont {Hiroki}\ \bibnamefont
  {Isobe}}\ and\ \bibinfo {author} {\bibfnamefont {Liang}\ \bibnamefont {Fu}},\
  }\bibfield  {title} {\enquote {\bibinfo {title} {Theory of interacting
  topological crystalline insulators},}\ }\href {\doibase
  10.1103/PhysRevB.92.081304} {\bibfield  {journal} {\bibinfo  {journal} {Phys.
  Rev. B}\ }\textbf {\bibinfo {volume} {92}},\ \bibinfo {pages} {081304--}
  (\bibinfo {year} {2015})}\BibitemShut {NoStop}%
\bibitem [{\citenamefont {Cheng}\ \emph {et~al.}(2016)\citenamefont {Cheng},
  \citenamefont {Zaletel}, \citenamefont {Barkeshli}, \citenamefont
  {Vishwanath},\ and\ \citenamefont {Bonderson}}]{Cheng:2016aa}%
  \BibitemOpen
  \bibfield  {author} {\bibinfo {author} {\bibfnamefont {Meng}\ \bibnamefont
  {Cheng}}, \bibinfo {author} {\bibfnamefont {Michael}\ \bibnamefont
  {Zaletel}}, \bibinfo {author} {\bibfnamefont {Maissam}\ \bibnamefont
  {Barkeshli}}, \bibinfo {author} {\bibfnamefont {Ashvin}\ \bibnamefont
  {Vishwanath}}, \ and\ \bibinfo {author} {\bibfnamefont {Parsa}\ \bibnamefont
  {Bonderson}},\ }\bibfield  {title} {\enquote {\bibinfo {title} {Translational
  symmetry and microscopic constraints on symmetry-enriched topological phases:
  A view from the surface},}\ }\href {\doibase 10.1103/PhysRevX.6.041068}
  {\bibfield  {journal} {\bibinfo  {journal} {Phys. Rev. X}\ }\textbf {\bibinfo
  {volume} {6}},\ \bibinfo {pages} {041068--} (\bibinfo {year}
  {2016})}\BibitemShut {NoStop}%
\bibitem [{\citenamefont {Chiu}\ \emph {et~al.}(2016)\citenamefont {Chiu},
  \citenamefont {Teo}, \citenamefont {Schnyder},\ and\ \citenamefont
  {Ryu}}]{Chiu:2016aa}%
  \BibitemOpen
  \bibfield  {author} {\bibinfo {author} {\bibfnamefont {Ching-Kai}\
  \bibnamefont {Chiu}}, \bibinfo {author} {\bibfnamefont {Jeffrey C.~Y.}\
  \bibnamefont {Teo}}, \bibinfo {author} {\bibfnamefont {Andreas~P.}\
  \bibnamefont {Schnyder}}, \ and\ \bibinfo {author} {\bibfnamefont {Shinsei}\
  \bibnamefont {Ryu}},\ }\bibfield  {title} {\enquote {\bibinfo {title}
  {Classification of topological quantum matter with symmetries},}\ }\href
  {https://link.aps.org/doi/10.1103/RevModPhys.88.035005} {\bibfield  {journal}
  {\bibinfo  {journal} {Rev. Mod. Phys.}\ }\textbf {\bibinfo {volume} {88}},\
  \bibinfo {pages} {035005--} (\bibinfo {year} {2016})}\BibitemShut {NoStop}%
\bibitem [{\citenamefont {Hermele}\ and\ \citenamefont
  {Chen}(2016)}]{Hermele:2016ws}%
  \BibitemOpen
  \bibfield  {author} {\bibinfo {author} {\bibfnamefont {Michael}\ \bibnamefont
  {Hermele}}\ and\ \bibinfo {author} {\bibfnamefont {Xie}\ \bibnamefont
  {Chen}},\ }\bibfield  {title} {\enquote {\bibinfo {title} {Flux-fusion
  anomaly test and bosonic topological crystalline insulators},}\ }\href
  {\doibase 10.1103/PhysRevX.6.041006} {\bibfield  {journal} {\bibinfo
  {journal} {Phys. Rev. X}\ }\textbf {\bibinfo {volume} {6}},\ \bibinfo {pages}
  {041006--} (\bibinfo {year} {2016})}\BibitemShut {NoStop}%
\bibitem [{\citenamefont {Thorngren}\ and\ \citenamefont
  {Else}(2018)}]{Thorngren:2018aa}%
  \BibitemOpen
  \bibfield  {author} {\bibinfo {author} {\bibfnamefont {Ryan}\ \bibnamefont
  {Thorngren}}\ and\ \bibinfo {author} {\bibfnamefont {Dominic~V.}\
  \bibnamefont {Else}},\ }\bibfield  {title} {\enquote {\bibinfo {title}
  {Gauging spatial symmetries and the classification of topological crystalline
  phases},}\ }\href {\doibase 10.1103/PhysRevX.8.011040} {\bibfield  {journal}
  {\bibinfo  {journal} {Phys. Rev. X}\ }\textbf {\bibinfo {volume} {8}},\
  \bibinfo {pages} {011040--} (\bibinfo {year} {2018})}\BibitemShut {NoStop}%
\bibitem [{\citenamefont {Else}\ and\ \citenamefont
  {Thorngren}(2019)}]{Else:2019ue}%
  \BibitemOpen
  \bibfield  {author} {\bibinfo {author} {\bibfnamefont {Dominic~V.}\
  \bibnamefont {Else}}\ and\ \bibinfo {author} {\bibfnamefont {Ryan}\
  \bibnamefont {Thorngren}},\ }\bibfield  {title} {\enquote {\bibinfo {title}
  {Crystalline topological phases as defect networks},}\ }\href {\doibase
  10.1103/PhysRevB.99.115116} {\bibfield  {journal} {\bibinfo  {journal} {Phys.
  Rev. B}\ }\textbf {\bibinfo {volume} {99}},\ \bibinfo {pages} {115116--}
  (\bibinfo {year} {2019})}\BibitemShut {NoStop}%
\bibitem [{\citenamefont {Lieb}\ \emph {et~al.}(1961)\citenamefont {Lieb},
  \citenamefont {Schultz},\ and\ \citenamefont {Mattis}}]{Lieb:1961aa}%
  \BibitemOpen
  \bibfield  {author} {\bibinfo {author} {\bibfnamefont {Elliott}\ \bibnamefont
  {Lieb}}, \bibinfo {author} {\bibfnamefont {Theodore}\ \bibnamefont
  {Schultz}}, \ and\ \bibinfo {author} {\bibfnamefont {Daniel}\ \bibnamefont
  {Mattis}},\ }\bibfield  {title} {\enquote {\bibinfo {title} {Two soluble
  models of an antiferromagnetic chain},}\ }\href
  {https://www.sciencedirect.com/science/article/abs/pii/0003491661901154}
  {\bibfield  {journal} {\bibinfo  {journal} {Ann. Phys.}\ }\textbf {\bibinfo
  {volume} {16}},\ \bibinfo {pages} {407--466} (\bibinfo {year}
  {1961})}\BibitemShut {NoStop}%
\bibitem [{\citenamefont {Affleck}\ and\ \citenamefont
  {Lieb}(1986)}]{Affleck:1986aa}%
  \BibitemOpen
  \bibfield  {author} {\bibinfo {author} {\bibfnamefont {Ian}\ \bibnamefont
  {Affleck}}\ and\ \bibinfo {author} {\bibfnamefont {Elliott~H}\ \bibnamefont
  {Lieb}},\ }\bibfield  {title} {\enquote {\bibinfo {title} {A proof of part of
  {H}aldane's conjecture on spin chains},}\ }\href
  {https://link.springer.com/article/10.1007/BF00400304} {\bibfield  {journal}
  {\bibinfo  {journal} {Lett. Math. Phys.}\ }\textbf {\bibinfo {volume} {12}},\
  \bibinfo {pages} {57--69} (\bibinfo {year} {1986})}\BibitemShut {NoStop}%
\bibitem [{\citenamefont {Oshikawa}(2000)}]{Oshikawa:2000aa}%
  \BibitemOpen
  \bibfield  {author} {\bibinfo {author} {\bibfnamefont {Masaki}\ \bibnamefont
  {Oshikawa}},\ }\bibfield  {title} {\enquote {\bibinfo {title}
  {Commensurability, excitation gap, and topology in quantum many-particle
  systems on a periodic lattice},}\ }\href
  {https://link.aps.org/doi/10.1103/PhysRevLett.84.1535} {\bibfield  {journal}
  {\bibinfo  {journal} {Phys. Rev. Lett.}\ }\textbf {\bibinfo {volume} {84}},\
  \bibinfo {pages} {1535--1538} (\bibinfo {year} {2000})}\BibitemShut {NoStop}%
\bibitem [{\citenamefont {Hastings}(2004)}]{Hastings:2004ab}%
  \BibitemOpen
  \bibfield  {author} {\bibinfo {author} {\bibfnamefont {M.~B.}\ \bibnamefont
  {Hastings}},\ }\bibfield  {title} {\enquote {\bibinfo {title}
  {Lieb-{S}chultz-{M}attis in higher dimensions},}\ }\href
  {https://link.aps.org/doi/10.1103/PhysRevB.69.104431} {\bibfield  {journal}
  {\bibinfo  {journal} {Phys. Rev. B}\ }\textbf {\bibinfo {volume} {69}},\
  \bibinfo {pages} {104431--} (\bibinfo {year} {2004})}\BibitemShut {NoStop}%
\bibitem [{\citenamefont {Po}\ \emph {et~al.}(2017)\citenamefont {Po},
  \citenamefont {Watanabe}, \citenamefont {Jian},\ and\ \citenamefont
  {Zaletel}}]{Po:2017aa}%
  \BibitemOpen
  \bibfield  {author} {\bibinfo {author} {\bibfnamefont {Hoi~Chun}\
  \bibnamefont {Po}}, \bibinfo {author} {\bibfnamefont {Haruki}\ \bibnamefont
  {Watanabe}}, \bibinfo {author} {\bibfnamefont {Chao-Ming}\ \bibnamefont
  {Jian}}, \ and\ \bibinfo {author} {\bibfnamefont {Michael~P.}\ \bibnamefont
  {Zaletel}},\ }\bibfield  {title} {\enquote {\bibinfo {title} {Lattice
  homotopy constraints on phases of quantum magnets},}\ }\href {\doibase
  10.1103/PhysRevLett.119.127202} {\bibfield  {journal} {\bibinfo  {journal}
  {Phys. Rev. Lett.}\ }\textbf {\bibinfo {volume} {119}},\ \bibinfo {pages}
  {127202--} (\bibinfo {year} {2017})}\BibitemShut {NoStop}%
\bibitem [{\citenamefont {Shiozaki}\ \emph {et~al.}(2022)\citenamefont
  {Shiozaki}, \citenamefont {Sato},\ and\ \citenamefont
  {Gomi}}]{Shiozaki:2022wm}%
  \BibitemOpen
  \bibfield  {author} {\bibinfo {author} {\bibfnamefont {Ken}\ \bibnamefont
  {Shiozaki}}, \bibinfo {author} {\bibfnamefont {Masatoshi}\ \bibnamefont
  {Sato}}, \ and\ \bibinfo {author} {\bibfnamefont {Kiyonori}\ \bibnamefont
  {Gomi}},\ }\bibfield  {title} {\enquote {\bibinfo {title} {Atiyah-hirzebruch
  spectral sequence in band topology: General formalism and topological
  invariants for 230 space groups},}\ }\href {\doibase
  10.1103/PhysRevB.106.165103} {\bibfield  {journal} {\bibinfo  {journal}
  {Phys. Rev. B}\ }\textbf {\bibinfo {volume} {106}},\ \bibinfo {pages}
  {165103--} (\bibinfo {year} {2022})}\BibitemShut {NoStop}%
\bibitem [{\citenamefont {Ogata}\ and\ \citenamefont
  {Tasaki}(2019)}]{Ogata:2018aa}%
  \BibitemOpen
  \bibfield  {author} {\bibinfo {author} {\bibfnamefont {Yoshiko}\ \bibnamefont
  {Ogata}}\ and\ \bibinfo {author} {\bibfnamefont {Hal}\ \bibnamefont
  {Tasaki}},\ }\bibfield  {title} {\enquote {\bibinfo {title}
  {Lieb--schultz--mattis type theorems for quantum spin chains without
  continuous symmetry},}\ }\href {\doibase 10.1007/s00220-019-03343-5}
  {\bibfield  {journal} {\bibinfo  {journal} {Comm. Math. Phys.}\ }\textbf
  {\bibinfo {volume} {372}},\ \bibinfo {pages} {951--962} (\bibinfo {year}
  {2019})}\BibitemShut {NoStop}%
\bibitem [{\citenamefont {Else}\ and\ \citenamefont
  {Thorngren}(2020)}]{Else:2020aa}%
  \BibitemOpen
  \bibfield  {author} {\bibinfo {author} {\bibfnamefont {Dominic~V.}\
  \bibnamefont {Else}}\ and\ \bibinfo {author} {\bibfnamefont {Ryan}\
  \bibnamefont {Thorngren}},\ }\bibfield  {title} {\enquote {\bibinfo {title}
  {Topological theory of lieb-schultz-mattis theorems in quantum spin
  systems},}\ }\href {\doibase 10.1103/PhysRevB.101.224437} {\bibfield
  {journal} {\bibinfo  {journal} {Phys. Rev. B}\ }\textbf {\bibinfo {volume}
  {101}},\ \bibinfo {pages} {224437--} (\bibinfo {year} {2020})}\BibitemShut
  {NoStop}%
\bibitem [{\citenamefont {Ogata}\ \emph {et~al.}(2021)\citenamefont {Ogata},
  \citenamefont {Tachikawa},\ and\ \citenamefont {Tasaki}}]{Ogata:2020aa}%
  \BibitemOpen
  \bibfield  {author} {\bibinfo {author} {\bibfnamefont {Yoshiko}\ \bibnamefont
  {Ogata}}, \bibinfo {author} {\bibfnamefont {Yuji}\ \bibnamefont {Tachikawa}},
  \ and\ \bibinfo {author} {\bibfnamefont {Hal}\ \bibnamefont {Tasaki}},\
  }\bibfield  {title} {\enquote {\bibinfo {title} {General
  lieb--schultz--mattis type theorems for quantum spin chains},}\ }\href
  {\doibase 10.1007/s00220-021-04116-9} {\bibfield  {journal} {\bibinfo
  {journal} {Comm. Math. Phys.}\ }\textbf {\bibinfo {volume} {385}},\ \bibinfo
  {pages} {79--99} (\bibinfo {year} {2021})}\BibitemShut {NoStop}%
\bibitem [{\citenamefont {Furuya}\ and\ \citenamefont
  {Oshikawa}(2017)}]{Furuya:2017aa}%
  \BibitemOpen
  \bibfield  {author} {\bibinfo {author} {\bibfnamefont {Shunsuke~C.}\
  \bibnamefont {Furuya}}\ and\ \bibinfo {author} {\bibfnamefont {Masaki}\
  \bibnamefont {Oshikawa}},\ }\bibfield  {title} {\enquote {\bibinfo {title}
  {Symmetry protection of critical phases and a global anomaly in $1+1$
  dimensions},}\ }\href
  {https://link.aps.org/doi/10.1103/PhysRevLett.118.021601} {\bibfield
  {journal} {\bibinfo  {journal} {Phys. Rev. Lett.}\ }\textbf {\bibinfo
  {volume} {118}},\ \bibinfo {pages} {021601--} (\bibinfo {year}
  {2017})}\BibitemShut {NoStop}%
\bibitem [{\citenamefont {Yao}\ \emph {et~al.}(2019)\citenamefont {Yao},
  \citenamefont {Hsieh},\ and\ \citenamefont {Oshikawa}}]{Yao:2019aa}%
  \BibitemOpen
  \bibfield  {author} {\bibinfo {author} {\bibfnamefont {Yuan}\ \bibnamefont
  {Yao}}, \bibinfo {author} {\bibfnamefont {Chang-Tse}\ \bibnamefont {Hsieh}},
  \ and\ \bibinfo {author} {\bibfnamefont {Masaki}\ \bibnamefont {Oshikawa}},\
  }\bibfield  {title} {\enquote {\bibinfo {title} {Anomaly matching and
  symmetry-protected critical phases in {\$}su(n){\$} spin systems in
  {\$}1+1{\$} dimensions},}\ }\href {\doibase 10.1103/PhysRevLett.123.180201}
  {\bibfield  {journal} {\bibinfo  {journal} {Phys. Rev. Lett.}\ }\textbf
  {\bibinfo {volume} {123}},\ \bibinfo {pages} {180201--} (\bibinfo {year}
  {2019})}\BibitemShut {NoStop}%
\bibitem [{\citenamefont {Lu}\ \emph {et~al.}(2020)\citenamefont {Lu},
  \citenamefont {Ran},\ and\ \citenamefont {Oshikawa}}]{Lu:2017aa}%
  \BibitemOpen
  \bibfield  {author} {\bibinfo {author} {\bibfnamefont {Yuan-Ming}\
  \bibnamefont {Lu}}, \bibinfo {author} {\bibfnamefont {Ying}\ \bibnamefont
  {Ran}}, \ and\ \bibinfo {author} {\bibfnamefont {Masaki}\ \bibnamefont
  {Oshikawa}},\ }\bibfield  {title} {\enquote {\bibinfo {title}
  {Filling-enforced constraint on the quantized hall conductivity on a periodic
  lattice},}\ }\href {\doibase https://doi.org/10.1016/j.aop.2019.168060}
  {\bibfield  {journal} {\bibinfo  {journal} {Ann. Phys.}\ }\textbf {\bibinfo
  {volume} {413}},\ \bibinfo {pages} {168060} (\bibinfo {year}
  {2020})}\BibitemShut {NoStop}%
\bibitem [{\citenamefont {Yao}\ and\ \citenamefont
  {Oshikawa}(2021)}]{Yao:2021aa}%
  \BibitemOpen
  \bibfield  {author} {\bibinfo {author} {\bibfnamefont {Yuan}\ \bibnamefont
  {Yao}}\ and\ \bibinfo {author} {\bibfnamefont {Masaki}\ \bibnamefont
  {Oshikawa}},\ }\bibfield  {title} {\enquote {\bibinfo {title} {Twisted
  boundary condition and lieb-schultz-mattis ingappability for discrete
  symmetries},}\ }\href {\doibase 10.1103/PhysRevLett.126.217201} {\bibfield
  {journal} {\bibinfo  {journal} {Phys. Rev. Lett.}\ }\textbf {\bibinfo
  {volume} {126}},\ \bibinfo {pages} {217201--} (\bibinfo {year}
  {2021})}\BibitemShut {NoStop}%
\bibitem [{\citenamefont {Shirley}\ \emph {et~al.}(2018)\citenamefont
  {Shirley}, \citenamefont {Slagle}, \citenamefont {Wang},\ and\ \citenamefont
  {Chen}}]{Shirley:2018tg}%
  \BibitemOpen
  \bibfield  {author} {\bibinfo {author} {\bibfnamefont {Wilbur}\ \bibnamefont
  {Shirley}}, \bibinfo {author} {\bibfnamefont {Kevin}\ \bibnamefont {Slagle}},
  \bibinfo {author} {\bibfnamefont {Zhenghan}\ \bibnamefont {Wang}}, \ and\
  \bibinfo {author} {\bibfnamefont {Xie}\ \bibnamefont {Chen}},\ }\bibfield
  {title} {\enquote {\bibinfo {title} {Fracton models on general
  three-dimensional manifolds},}\ }\href {\doibase 10.1103/PhysRevX.8.031051}
  {\bibfield  {journal} {\bibinfo  {journal} {Phys. Rev. X}\ }\textbf {\bibinfo
  {volume} {8}},\ \bibinfo {pages} {031051--} (\bibinfo {year}
  {2018})}\BibitemShut {NoStop}%
\bibitem [{\citenamefont {Fu}\ \emph {et~al.}(2007)\citenamefont {Fu},
  \citenamefont {Kane},\ and\ \citenamefont {Mele}}]{Fu:2007aa}%
  \BibitemOpen
  \bibfield  {author} {\bibinfo {author} {\bibfnamefont {Liang}\ \bibnamefont
  {Fu}}, \bibinfo {author} {\bibfnamefont {C.~L.}\ \bibnamefont {Kane}}, \ and\
  \bibinfo {author} {\bibfnamefont {E.~J.}\ \bibnamefont {Mele}},\ }\bibfield
  {title} {\enquote {\bibinfo {title} {Topological insulators in three
  dimensions},}\ }\href {\doibase 10.1103/PhysRevLett.98.106803} {\bibfield
  {journal} {\bibinfo  {journal} {Phys. Rev. Lett.}\ }\textbf {\bibinfo
  {volume} {98}},\ \bibinfo {pages} {106803--} (\bibinfo {year}
  {2007})}\BibitemShut {NoStop}%
\bibitem [{\citenamefont {Witten}(2016{\natexlab{b}})}]{Witten:2016ab}%
  \BibitemOpen
  \bibfield  {author} {\bibinfo {author} {\bibfnamefont {Edward}\ \bibnamefont
  {Witten}},\ }\bibfield  {title} {\enquote {\bibinfo {title} {The ``parity''
  anomaly on an unorientable manifold},}\ }\href {\doibase
  10.1103/PhysRevB.94.195150} {\bibfield  {journal} {\bibinfo  {journal} {Phys.
  Rev. B}\ }\textbf {\bibinfo {volume} {94}},\ \bibinfo {pages} {195150--}
  (\bibinfo {year} {2016}{\natexlab{b}})}\BibitemShut {NoStop}%
\bibitem [{\citenamefont {Wang}\ \emph {et~al.}(2018)\citenamefont {Wang},
  \citenamefont {Wen},\ and\ \citenamefont {Witten}}]{Wang:2018um}%
  \BibitemOpen
  \bibfield  {author} {\bibinfo {author} {\bibfnamefont {Juven}\ \bibnamefont
  {Wang}}, \bibinfo {author} {\bibfnamefont {Xiao-Gang}\ \bibnamefont {Wen}}, \
  and\ \bibinfo {author} {\bibfnamefont {Edward}\ \bibnamefont {Witten}},\
  }\bibfield  {title} {\enquote {\bibinfo {title} {Symmetric gapped interfaces
  of spt and set states: Systematic constructions},}\ }\href {\doibase
  10.1103/PhysRevX.8.031048} {\bibfield  {journal} {\bibinfo  {journal} {Phys.
  Rev. X}\ }\textbf {\bibinfo {volume} {8}},\ \bibinfo {pages} {031048--}
  (\bibinfo {year} {2018})}\BibitemShut {NoStop}%
\bibitem [{\citenamefont {Chen}\ \emph {et~al.}(2011)\citenamefont {Chen},
  \citenamefont {Gu},\ and\ \citenamefont
  {Wen}}]{Chen-Gu-Wen_classification2010}%
  \BibitemOpen
  \bibfield  {author} {\bibinfo {author} {\bibfnamefont {Xie}\ \bibnamefont
  {Chen}}, \bibinfo {author} {\bibfnamefont {Zheng-Cheng}\ \bibnamefont {Gu}},
  \ and\ \bibinfo {author} {\bibfnamefont {Xiao-Gang}\ \bibnamefont {Wen}},\
  }\bibfield  {title} {\enquote {\bibinfo {title} {Classification of gapped
  symmetric phases in one-dimensional spin systems},}\ }\href {\doibase
  10.1103/PhysRevB.83.035107} {\bibfield  {journal} {\bibinfo  {journal} {Phys.
  Rev. B}\ }\textbf {\bibinfo {volume} {83}},\ \bibinfo {pages} {035107}
  (\bibinfo {year} {2011})}\BibitemShut {NoStop}%
\bibitem [{\citenamefont {Chen}\ \emph {et~al.}(2013)\citenamefont {Chen},
  \citenamefont {Gu}, \citenamefont {Liu},\ and\ \citenamefont
  {Wen}}]{Chen:2013aa}%
  \BibitemOpen
  \bibfield  {author} {\bibinfo {author} {\bibfnamefont {Xie}\ \bibnamefont
  {Chen}}, \bibinfo {author} {\bibfnamefont {Zheng-Cheng}\ \bibnamefont {Gu}},
  \bibinfo {author} {\bibfnamefont {Zheng-Xin}\ \bibnamefont {Liu}}, \ and\
  \bibinfo {author} {\bibfnamefont {Xiao-Gang}\ \bibnamefont {Wen}},\
  }\bibfield  {title} {\enquote {\bibinfo {title} {Symmetry protected
  topological orders and the group cohomology of their symmetry group},}\
  }\href {\doibase 10.1103/PhysRevB.87.155114} {\bibfield  {journal} {\bibinfo
  {journal} {Phys. Rev. B}\ }\textbf {\bibinfo {volume} {87}},\ \bibinfo
  {pages} {155114--} (\bibinfo {year} {2013})}\BibitemShut {NoStop}%
\bibitem [{\citenamefont {Xiong}(2018)}]{Xiong:2018aa}%
  \BibitemOpen
  \bibfield  {author} {\bibinfo {author} {\bibfnamefont {Charles~Zhaoxi}\
  \bibnamefont {Xiong}},\ }\bibfield  {title} {\enquote {\bibinfo {title}
  {Minimalist approach to the classification of symmetry protected topological
  phases},}\ }\href@noop {} {\bibfield  {journal} {\bibinfo  {journal} {J.
  Phys. A: Math. Theor.}\ }\textbf {\bibinfo {volume} {51}},\ \bibinfo {pages}
  {445001} (\bibinfo {year} {2018})}\BibitemShut {NoStop}%
\bibitem [{\citenamefont {Aasen}\ \emph {et~al.}(2022)\citenamefont {Aasen},
  \citenamefont {Wang},\ and\ \citenamefont {Hastings}}]{Aasen:2022aa}%
  \BibitemOpen
  \bibfield  {author} {\bibinfo {author} {\bibfnamefont {David}\ \bibnamefont
  {Aasen}}, \bibinfo {author} {\bibfnamefont {Zhenghan}\ \bibnamefont {Wang}},
  \ and\ \bibinfo {author} {\bibfnamefont {Matthew~B.}\ \bibnamefont
  {Hastings}},\ }\bibfield  {title} {\enquote {\bibinfo {title} {Adiabatic
  paths of hamiltonians, symmetries of topological order, and automorphism
  codes},}\ }\href {\doibase 10.1103/PhysRevB.106.085122} {\bibfield  {journal}
  {\bibinfo  {journal} {Phys. Rev. B}\ }\textbf {\bibinfo {volume} {106}},\
  \bibinfo {pages} {085122--} (\bibinfo {year} {2022})}\BibitemShut {NoStop}%
\bibitem [{\citenamefont {Haldane}(1988)}]{Haldane:1988uf}%
  \BibitemOpen
  \bibfield  {author} {\bibinfo {author} {\bibfnamefont {F.~D.~M.}\
  \bibnamefont {Haldane}},\ }\bibfield  {title} {\enquote {\bibinfo {title}
  {Model for a quantum hall effect without landau levels: Condensed-matter
  realization of the "parity anomaly"},}\ }\href {\doibase
  10.1103/PhysRevLett.61.2015} {\bibfield  {journal} {\bibinfo  {journal}
  {Phys. Rev. Lett.}\ }\textbf {\bibinfo {volume} {61}},\ \bibinfo {pages}
  {2015--2018} (\bibinfo {year} {1988})}\BibitemShut {NoStop}%
\bibitem [{\citenamefont {Laughlin}(1981)}]{Laughlin:1981aa}%
  \BibitemOpen
  \bibfield  {author} {\bibinfo {author} {\bibfnamefont {R.~B.}\ \bibnamefont
  {Laughlin}},\ }\bibfield  {title} {\enquote {\bibinfo {title} {Quantized hall
  conductivity in two dimensions},}\ }\href {\doibase 10.1103/PhysRevB.23.5632}
  {\bibfield  {journal} {\bibinfo  {journal} {Phys. Rev. B}\ }\textbf {\bibinfo
  {volume} {23}},\ \bibinfo {pages} {5632--5633} (\bibinfo {year}
  {1981})}\BibitemShut {NoStop}%
\bibitem [{\citenamefont {Lu}\ and\ \citenamefont
  {Vishwanath}(2012)}]{Lu:2012ux}%
  \BibitemOpen
  \bibfield  {author} {\bibinfo {author} {\bibfnamefont {Yuan-Ming}\
  \bibnamefont {Lu}}\ and\ \bibinfo {author} {\bibfnamefont {Ashvin}\
  \bibnamefont {Vishwanath}},\ }\bibfield  {title} {\enquote {\bibinfo {title}
  {Theory and classification of interacting integer topological phases in two
  dimensions: A chern-simons approach},}\ }\href {\doibase
  10.1103/PhysRevB.86.125119} {\bibfield  {journal} {\bibinfo  {journal} {Phys.
  Rev. B}\ }\textbf {\bibinfo {volume} {86}},\ \bibinfo {pages} {125119--}
  (\bibinfo {year} {2012})}\BibitemShut {NoStop}%
\bibitem [{\citenamefont {Wang}(2015)}]{Wang:2015vv}%
  \BibitemOpen
  \bibfield  {author} {\bibinfo {author} {\bibfnamefont {Chong}\ \bibnamefont
  {Wang}},\ }\bibfield  {title} {\enquote {\bibinfo {title} {Bound states of
  three fermions forming symmetry-protected topological phases},}\ }\href
  {\doibase 10.1103/PhysRevB.91.245124} {\bibfield  {journal} {\bibinfo
  {journal} {Phys. Rev. B}\ }\textbf {\bibinfo {volume} {91}},\ \bibinfo
  {pages} {245124--} (\bibinfo {year} {2015})}\BibitemShut {NoStop}%
\bibitem [{\citenamefont {Senthil}\ and\ \citenamefont
  {Levin}(2013)}]{Senthil:2013tp}%
  \BibitemOpen
  \bibfield  {author} {\bibinfo {author} {\bibfnamefont {T.}~\bibnamefont
  {Senthil}}\ and\ \bibinfo {author} {\bibfnamefont {Michael}\ \bibnamefont
  {Levin}},\ }\bibfield  {title} {\enquote {\bibinfo {title} {Integer quantum
  hall effect for bosons},}\ }\href {\doibase 10.1103/PhysRevLett.110.046801}
  {\bibfield  {journal} {\bibinfo  {journal} {Phys. Rev. Lett.}\ }\textbf
  {\bibinfo {volume} {110}},\ \bibinfo {pages} {046801--} (\bibinfo {year}
  {2013})}\BibitemShut {NoStop}%
\bibitem [{\citenamefont {Yao}\ \emph {et~al.}()\citenamefont {Yao},
  \citenamefont {Li}, \citenamefont {Hsieh},\ and\ \citenamefont
  {Oshikawa}}]{YaoLiHsiehOshikawa}%
  \BibitemOpen
  \bibfield  {author} {\bibinfo {author} {\bibfnamefont {Y.}~\bibnamefont
  {Yao}}, \bibinfo {author} {\bibfnamefont {L.}~\bibnamefont {Li}}, \bibinfo
  {author} {\bibfnamefont {C.T.}\ \bibnamefont {Hsieh}}, \ and\ \bibinfo
  {author} {\bibfnamefont {M.}~\bibnamefont {Oshikawa}},\ }\href@noop {}
  {\bibinfo  {journal} {unpublished}\ }\BibitemShut {NoStop}%
\bibitem [{\citenamefont {Read}\ and\ \citenamefont
  {Green}(2000)}]{Read:2000vz}%
  \BibitemOpen
\bibfield  {journal} {  }\bibfield  {author} {\bibinfo {author} {\bibfnamefont
  {N.}~\bibnamefont {Read}}\ and\ \bibinfo {author} {\bibfnamefont {Dmitry}\
  \bibnamefont {Green}},\ }\bibfield  {title} {\enquote {\bibinfo {title}
  {Paired states of fermions in two dimensions with breaking of parity and
  time-reversal symmetries and the fractional quantum hall effect},}\ }\href
  {\doibase 10.1103/PhysRevB.61.10267} {\bibfield  {journal} {\bibinfo
  {journal} {Phys. Rev. B}\ }\textbf {\bibinfo {volume} {61}},\ \bibinfo
  {pages} {10267--10297} (\bibinfo {year} {2000})}\BibitemShut {NoStop}%
\bibitem [{\citenamefont {Marshall}(1955)}]{Marshall:1955tw}%
  \BibitemOpen
  \bibfield  {author} {\bibinfo {author} {\bibfnamefont {W}~\bibnamefont
  {Marshall}},\ }\bibfield  {title} {\enquote {\bibinfo {title}
  {Antiferromagnetism},}\ }\href
  {https://royalsocietypublishing.org/doi/10.1098/rspa.1955.0200} {\bibfield
  {journal} {\bibinfo  {journal} {Proc. R. Soc. A (London)}\ }\textbf {\bibinfo
  {volume} {232}},\ \bibinfo {pages} {48--68} (\bibinfo {year}
  {1955})}\BibitemShut {NoStop}%
\bibitem [{\citenamefont {Lieb}\ and\ \citenamefont
  {Mattis}(1962)}]{Lieb:1962wk}%
  \BibitemOpen
  \bibfield  {author} {\bibinfo {author} {\bibfnamefont {Elliott}\ \bibnamefont
  {Lieb}}\ and\ \bibinfo {author} {\bibfnamefont {Daniel}\ \bibnamefont
  {Mattis}},\ }\bibfield  {title} {\enquote {\bibinfo {title} {Ordering energy
  levels of interacting spin systems},}\ }\href
  {https://link.springer.com/content/pdf/10.1007/978-3-642-55925-9_3.pdf}
  {\bibfield  {journal} {\bibinfo  {journal} {J. Math. Phys.}\ }\textbf
  {\bibinfo {volume} {3}},\ \bibinfo {pages} {749--751} (\bibinfo {year}
  {1962})}\BibitemShut {NoStop}%
\bibitem [{\citenamefont {Lieb}(1989)}]{Lieb:1989wr}%
  \BibitemOpen
  \bibfield  {author} {\bibinfo {author} {\bibfnamefont {Elliott~H.}\
  \bibnamefont {Lieb}},\ }\bibfield  {title} {\enquote {\bibinfo {title} {Two
  theorems on the hubbard model},}\ }\href {\doibase
  10.1103/PhysRevLett.62.1201} {\bibfield  {journal} {\bibinfo  {journal}
  {Phys. Rev. Lett.}\ }\textbf {\bibinfo {volume} {62}},\ \bibinfo {pages}
  {1201--1204} (\bibinfo {year} {1989})}\BibitemShut {NoStop}%
\bibitem [{\citenamefont {Chitov}\ \emph {et~al.}(2008)\citenamefont {Chitov},
  \citenamefont {Ramakko},\ and\ \citenamefont {Azzouz}}]{Chitov:2008aa}%
  \BibitemOpen
  \bibfield  {author} {\bibinfo {author} {\bibfnamefont {Gennady~Y}\
  \bibnamefont {Chitov}}, \bibinfo {author} {\bibfnamefont {Brandon~W}\
  \bibnamefont {Ramakko}}, \ and\ \bibinfo {author} {\bibfnamefont {Mohamed}\
  \bibnamefont {Azzouz}},\ }\bibfield  {title} {\enquote {\bibinfo {title}
  {Quantum criticality in dimerized spin ladders},}\ }\href {\doibase
  10.1103/PhysRevB.77.224433} {\bibfield  {journal} {\bibinfo  {journal} {Phys.
  Rev. B}\ }\textbf {\bibinfo {volume} {77}},\ \bibinfo {pages} {224433}
  (\bibinfo {year} {2008})}\BibitemShut {NoStop}%
\end{thebibliography}

%

\end{document}